\documentclass[prd,showpacs,showkeys,preprintnumbers,aps,amssymb]{revtex4}

\usepackage{amsmath,amssymb}
\usepackage{pstricks}
\usepackage{graphicx}
\usepackage{dcolumn}
\usepackage{bm}

\newcommand{\tr}{\textrm{tr}}
\renewcommand{\theequation}{\arabic{section}.\arabic{equation}}
\def\phs#1{%
\scalebox{0.8}{\parbox{2.5cm}{\includegraphics[keepaspectratio,height=1.3cm]{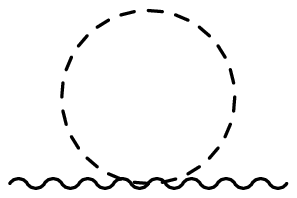}
\rput(-1.1,1.){\scalebox{1.}{#1}}} 
}}

\def\phss#1#2{
\scalebox{0.8}{ \parbox{3.cm}{\includegraphics[keepaspectratio,height=1.3cm]{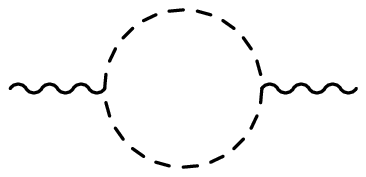}
\rput(-1.5,1.){\scalebox{1.}{#1}}
\rput(-1.5,.3){\scalebox{1.}{#2}}}
}}

\def\phvs#1#2{
\scalebox{0.8}{\parbox{3.cm}{\includegraphics[keepaspectratio,height=1.3cm]{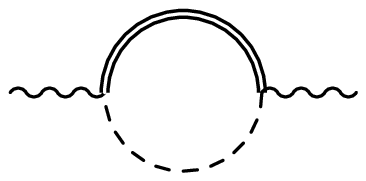}
\rput(-1.5,1.){\scalebox{1.}{#1}}\rput(-1.5,.3){\scalebox{1.}{#2}}}
}}

\def\scs#1{%
\scalebox{0.8}{\parbox{3.cm}{\includegraphics[keepaspectratio,height=1.3cm]{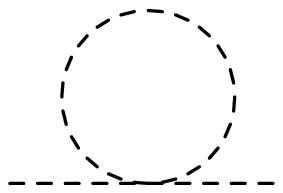}
\rput(-1.,1.0){\scalebox{1.}{#1}}} 
}}

\def\scv#1{%
\scalebox{0.8}{\parbox{3.cm}{\includegraphics[keepaspectratio,height=1.3cm]{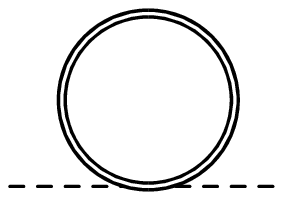}
\rput(-1.,1.0){\scalebox{1.}{#1}}} 
}}

\def\scss#1#2{
\scalebox{0.8}{ \parbox{3.cm}{\includegraphics[keepaspectratio,height=1.3cm]{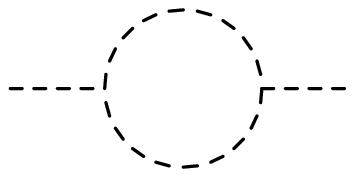}
\rput(-1.4,1.){\scalebox{1.}{#1}}\rput(-1.4,.3){\scalebox{1.}{#2}}}
}}

\def\scvs#1#2{
\scalebox{0.8}{\parbox{3.cm}{\includegraphics[keepaspectratio,height=1.3cm]{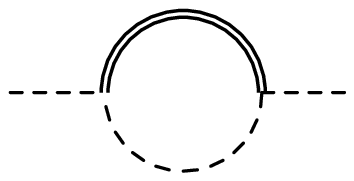}
\rput(-1.4,1.){\scalebox{1.}{#1}}\rput(-1.4,.3){\scalebox{1.}{#2}}}
}}

\def\scvv#1#2{
\scalebox{0.8}{\parbox{3.cm}{\includegraphics[keepaspectratio,height=1.3cm]{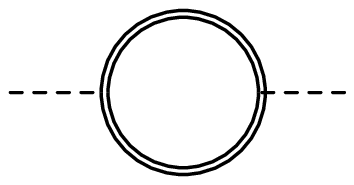}
\rput(-1.4,1.){\scalebox{1.}{#1}}\rput(-1.4,.3){\scalebox{1.}{#2}}}
}}

\def\vvs#1{%
\scalebox{0.8}{\parbox{3.cm}{\includegraphics[keepaspectratio,height=1.3cm]{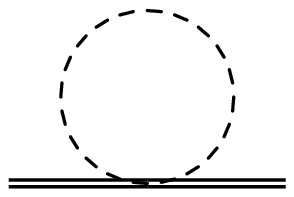}
\rput(-1.,1.){\scalebox{1.}{#1}}} 
}}

\def\vvv#1{%
\scalebox{0.8}{\parbox{3.cm}{\includegraphics[keepaspectratio,height=1.3cm]{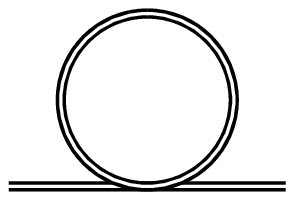}
\rput(-1.,1.){\scalebox{1.}{#1}}} 
}}

\def\vvss#1#2{
\scalebox{0.8}{ \parbox{3.cm}{\includegraphics[keepaspectratio,height=1.3cm]{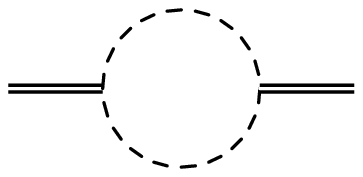}
\rput(-1.5,1.){\scalebox{1.}{#1}}\rput(-1.5,.3){\scalebox{1.}{#2}}}
}}

\def\vvvs#1#2{
\scalebox{0.8}{\parbox{3.cm}{\includegraphics[keepaspectratio,height=1.3cm]{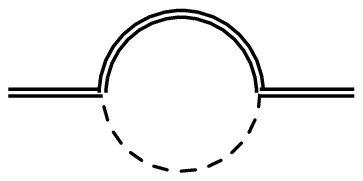}
\rput(-1.5,1.){\scalebox{1.}{#1}}\rput(-1.5,.3){\scalebox{1.}{#2}}}
}}

\def\vvvv#1#2{
\scalebox{0.8}{\parbox{3.cm}{\includegraphics[keepaspectratio,height=1.3cm]{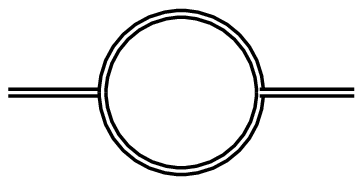}
\rput(-1.5,1.){\scalebox{1.}{#1}}\rput(-1.5,.3){\scalebox{1.}{#2}}}
}}

\def\vvgg#1#2{
\scalebox{0.8}{\parbox{3.cm}{\includegraphics[keepaspectratio,height=1.3cm]{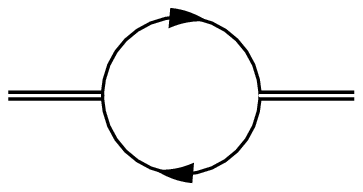}
\rput(-1.3,1.){\scalebox{1.}{#1}}\rput(-1.3,.3){\scalebox{1.}{#2}}}
}}

\def\pvs#1{%
\scalebox{0.8}{\parbox{2.7cm}{\includegraphics[keepaspectratio,height=1.3cm]{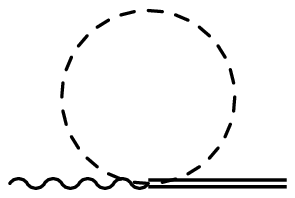}
\rput(-1.1,1.){\scalebox{1.}{#1}}} 
}}

\def\pvss#1#2{
\scalebox{0.8}{ \parbox{3.cm}{\includegraphics[keepaspectratio,height=1.3cm]{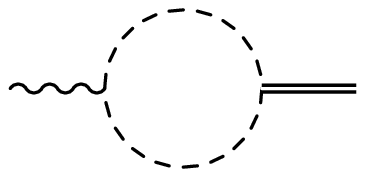}
\rput(-1.5,1.){\scalebox{1.}{#1}}\rput(-1.5,.3){\scalebox{1.}{#2}}}
}}

\def\pvvs#1#2{
\scalebox{0.8}{\parbox{3.cm}{\includegraphics[keepaspectratio,height=1.3cm]{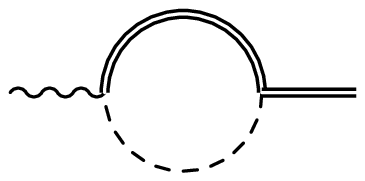}
\rput(-1.5,1.){\scalebox{1.}{#1}}\rput(-1.5,.3){\scalebox{1.}{#2}}}
}}

\begin{document}

\preprint{KEK-TH-1061}

\title{Renormalization group equations in a model of generalized hidden local 
symmetry and restoration of chiral symmetry}

\author{Yoshimasa Hidaka$^{\rm a}$%
\footnote{$\!\!$Electronic address: \tt hidaka@post.kek.jp},
Osamu Morimatsu$^{\rm a}$%
\footnote{$\!\!$Electronic address: \tt osamu.morimatsu@kek.jp}
 and Munehisa Ohtani$^{\rm b}$%
\footnote{$\!\!$Electronic address: \tt ohtani@rarfaxp.riken.jp}} 
\affiliation{%
$^{\rm a}$Institute of Particle and Nuclear Studies, High Energy Accelerator Research Organization, 1-1, Ooho, Tsukuba, Ibaraki, 305-0801, Japan \\
$^{\rm b}$Radiation Laboratory, RIKEN, Wako, Saitama, 351-0198, Japan
}

\date{December 30, 2005}
\begin{abstract}
We study possible restoration patterns of chiral symmetry in a generalized hidden local symmetry model, which is a low energy effective theory of QCD including pseudo-scalar, vector and axial-vector mesons.
We derive Wilsonian renormalization group equations
and analyze the running couplings and their fixed points
at the chiral restoration point.
We find three types of the chiral restoration, 
which are classified as the standard,
vector manifestation and intermediate scenarios, respectively.
It turns out that the rho and $A_1$ meson become massless 
and their decay into pion is suppressed in all the restoration patterns.
The each restoration scenario violates or fulfills the vector meson dominance 
at the critical point in a different manner, which may
reflect on the contributions from the pion to the dilepton spectrum.
\end{abstract}

\pacs{11.10.Hi, 11.30.Rd, 12.39.Fe, 14.40.-n}
\keywords{rho meson, vector manifestation, hidden local symmetry, 
renormalization group, chiral symmetry}
\maketitle
\section{Introduction}
Dynamical chiral symmetry breaking is the most remarkable 
nonperturbative phenomenon in QCD together with confinement.
One of the most important goals in hot and/or dense QCD is to understand how chiral symmetry restores at high temperature and/or density, which is commonly believed to be true.
Then, it is crucially important to understand how hadrons change their properties in hot and/or dense matter, since it is what one observes in the actual experiments.
Along these lines there have already been many theoretical as well as 
experimental works \cite{Hatsuda:1994pi,Rapp:1999ej,Brown:2001nh, 
Agakichiev:1995xb,Bonutti:1996ij,Ozawa:2000iw, Scomparin:2005}.

Theoretically, the properties of pseudo-scalar mesons are determined by the chiral symmetry through low-energy theorems since pseudo-scalar mesons are considered to be approximate Goldstone bosons, while the properties of other mesons are not directly controlled by the symmetry alone.
An interesting possibility, the vector manifestation (VM), has recently 
been proposed by Harada and Yamawaki \cite{Harada:2001}, in which the vector meson becomes the chiral partner of the pseudo-scalar meson when the chiral symmetry is restored and provides a theoretical basis for Brown-Rho scaling \cite{Brown:1991kk}, i.e. the dropping of the vector-meson mass in hot and/or dense matter.
The VM is based on the hidden local symmetry (HLS) by Bando et al.\ 
\cite{Bando:1984ej,Bando:1984pw,Fujiwara:1984mp,Bando:1985rf,Bando:1988}, 
which is a symmetry not manifest in the QCD lagrangian but is assumed to be dynamically generated, and vector mesons are introduced as gauge bosons associated with the HLS.
Though there had been a long history before the work of  Bando et al.\ 
\cite{Yang:1954ek,Sakurai:1969,Schwinger:1967tc,Wess:1967jq,Weinberg:1968de,Gasser:1983yg}, in trying to incorporate vector mesons as gauge bosons in the low-energy effective field theory, the HLS made it possible to include vector mesons in a systematic loop expansion of the chiral perturbation theory \cite{Tanabashi:1993sr}.

In the broken phase of the chiral symmetry the pseudo-scalar and 
axial-vector mesons are mixed with each other while scalar and vector 
mesons are pure, when mesons are classified according to the 
representation associated with light-front chiral charges \cite{Gilman:1960, Weinberg:1969, Weinberg:1990xn,Yamawaki:1998cy}.
This mixing is expected to vanish in the symmetric phase and the mesons form chiral doublets.
It is usually believed that the scalar meson becomes the chiral partner of the pseudo-scalar meson.
This is called the \lq standard scenario\rq.
There is, however, another possibility that the longitudinal part of the vector meson becomes the chiral partner of the pseudo-scalar meson.
This is the \lq VM scenario\rq \ mentioned above.
Therefore, it is an interesting question, if the VM scenario is realized in the real world or not.
The VM is claimed to be realized in the HLS model at large number of flavors in ref.\cite{Harada:1999zj,Harada:2001b,Harada:2001} by Harada and Yamawaki, at high temperature in ref.\cite{Harada:2001it} by Harada and Sasaki and at high density by Harada, Kim and Rho in ref.\cite{Harada:2001qz}\ . 
There argument goes as follows.
The parameters of the HLS are determined from QCD by the Wilsonian 
matching proposed in ref.\cite{Harada:2001b}, namely by matching the vector and axial-vector current correlators in the HLS model with those in the operator product expansion (OPE) of QCD at the matching scale, $\Lambda$.
Then, they obtain the parameters of the HLS at lower physical energy by 
letting the parameters run from the scale, $\Lambda$, according to the renormalization group equation (RGE).
Now, when the chiral symmetry is restored, the vector and axial-vector current correlators must coincide with each other.
When the parameters of the HLS run according to the renormalization group equation, this condition is always satisfied while the pion decay constant, which is non-zero at the matching scale, $\Lambda$, goes to zero at lower energy, i.e. on-shell.
In the HLS model, the condition that  the vector and axial-vector current correlators coincide together with the vanishing pion decay constant leads to the VM, because 
only the pseudo-scalar and vector mesons are included in the model \cite{Harada:2001}.
Thus, it was concluded that the VM is realized in the symmetric phase.

These works, however,  are not fully satisfactory in a sense that the model of the HLS includes only the pseudo-scalar and vector mesons but neither scalar nor axial-vector mesons.
Namely, from the very beginning the possibility of realizing the standard scenario seems to be excluded.
Thus, a natural question is what happens if one includes the axial-vector and scalar mesons into consideration.
The purpose of this paper is to give a partial answer to the question by 
employing the generalized hidden local symmetry (GHLS) model \cite{Bando:1985rf,Bando:1987br}, which includes the axial-vector meson in addition to the pseudo-scalar and vector mesons.
We analyze the renormalization group equation in the GHLS model and investigate which scenario is realized in the chiral symmetric phase.
It should be noted that importance of the  axial-vector meson is 
stressed from the viewpoint of the large $\pi-A_1$ mixing and also of 
recent results of STAR collaboration \cite{STAR:2004,Brown:2005}.

The paper is organized as follows. The concept of GHLS is introduced in sec.\ \ref{sec:HLS}.  The Wilsonian matching for the GHLS model is performed in sec.\ \ref{sec:WilsonMatch}. The renormalization group equations and their fixed points are given in sec.\ \ref{sec:RGEs}. Possible restoration patterns are examined in sec.\ \ref{sec:Restpattern}. Sec.\ \ref{sec:summary} is devoted to the summary of the paper. 
The actual calculation to obtain the coefficients of the renormalization 
group equations is extremely complicated and only the results are shown in sec.\ \ref{sec:RGEs}. We provide the minimum necessary information in order to reproduce the results in Appendix, still which is very long.
\setcounter{equation}{0}
\section{Hidden Local Symmetry}
\label{sec:HLS}
The concept of the hidden local symmetry is that a symmetry, which is not manifest in the fundamental theory, is dynamically generated in the effective theory.
The QCD with $N_f$ massless quarks and its low-energy effective theory, the non-linear sigma model shares the same global symmetry, $G_\textrm{global}=\textrm{SU}(N_f)_\textrm{L}\times \textrm{SU}(N_f)_\textrm{R}$, under which the pion field $U$ is transformed as
\[
U \to U'=g_\textrm{L}Ug_\textrm{R}^\dagger,
\]
where $g_{\textrm{L},\textrm{R}}\in G_\textrm{global}$.
In a model of the generalized hidden local symmetry with 
$G_\textrm{local}=\textrm{SU}(N_f)_\textrm{L}\times \textrm{SU}(N_f)_\textrm{R}$, the vector and axial-vector mesons are introduced as gauge bosons associated with the hidden local symmetry \cite{Bando:1985rf,Bando:1987br}. 
$G_\textrm{global}\times G_\textrm{local}$ are spontaneously broken down 
to global symmetry $H_\textrm{global}=\textrm{SU}(N)_\textrm{V}$.
The vector and axial-vector mesons acquire masses through the Higgs mechanism with spontaneous breaking of the GHLS.
We define representatives $\xi_\textrm{{L},{R},{M}}$
 in the quotient space $G_\textrm{global}\times G_\textrm{local}/H_\textrm{global}$
such that $U=\xi_\textrm{L}^\dagger\xi_\textrm{M}\xi_\textrm{R}$, 
and hidden local gauge bosons, $V_{\textrm{L},\textrm{R}}$.
These fields,
$\xi_{\textrm{L},\textrm{R},\textrm{M}}$ and $V_{\textrm{L},\textrm{R}}$,
are transformed under $G_\textrm{global}\times G_\textrm{local}$ as
\begin{align}
\xi_\textrm{L,R}&\rightarrow \xi'_\textrm{L,R}=G_\textrm{L,R}(x)\xi_\textrm{L,R}(x)g_\textrm{L,R}^\dagger,\nonumber\\
\xi_\textrm{M} &\rightarrow \xi'_\textrm{M}=G_\textrm{L}(x)\xi_\textrm{M}(x)G_\textrm{R}^\dagger(x),\nonumber\\
V_{\textrm{L,R}\mu}&\rightarrow V'_{\textrm{L,R}\mu}=G_\textrm{L,R}(x)V_{\textrm{L,R}\mu}G_\textrm{L,R}^\dagger(x)-i\partial_\mu G_\textrm{L,R}(x)G_\textrm{L,R}^\dagger(x),
\end{align}
$G_\textrm{L,R}(x)\in G_\textrm{local}$. 
The covariant derivative for $\xi_\textrm{L,R,M}$ are defined as
\begin{align}
D_\mu\xi_\textrm{L}&=\partial_\mu\xi_\textrm{L}-iV_{\textrm{L}\mu}\xi_\textrm{L}
+i\xi_\textrm{L}{\cal L}_\mu,\nonumber\\
D_\mu\xi_\textrm{R}&=\partial_\mu\xi_\textrm{R}-iV_{\textrm{R}\mu}\xi_\textrm{R}+i\xi_\textrm{R}{\cal R}_\mu,\nonumber\\
D_\mu\xi_\textrm{M}&=\partial_\mu\xi_\textrm{M}-iV_{\textrm{L}\mu}\xi_\textrm{M}+i\xi_\textrm{M}V_{\textrm{R}\mu},
\end{align}
where ${\cal L}_\mu$ and ${\cal R}_\mu$ are external gauge fields
which are associated with the global symmetry $G_{\rm global}$.
The covariantized 1-forms,
\begin{align}
\widehat\alpha_{\textrm{L,R}\mu}&=\frac{1}{i}D_\mu\xi_\textrm{L,R}\xi_\textrm{L,R}^\dagger, \ \ 
\widehat\alpha_{\textrm{M}\mu}=\frac{1}{2i}D_\mu\xi_\textrm{M}\xi_\textrm{M}^\dagger,
\end{align}
and field strength,
\begin{align}
F^{(\textrm{L,R})}_{\mu\nu}=\partial_\mu V_{\textrm{L,R}\nu}-\partial_\nu V_{\textrm{L,R}\mu}-i[V_{\textrm{L,R}\mu},V_{\textrm{L,R}\nu}],
\end{align}
are mapped under the gauge transformations as 
\begin{align}
\widehat\alpha_{\textrm{L,R}\mu}&\to G_\textrm{L,R}(x)\widehat\alpha_{\textrm{L,R}\mu}G_\textrm{L,R}^\dagger(x),\nonumber\\
\widehat\alpha_{\textrm{M}\mu}&\to G_\textrm{L}(x)\widehat\alpha_{\textrm{M}\mu}G_\textrm{L}^\dagger(x),\nonumber\\
F_{\mu\nu}^{(\textrm{L,R})}&\to G_\textrm{L,R}(x)F_{\mu\nu}^{(\textrm{L,R})}G_\textrm{L,R}^\dagger(x).
\end{align}
From these field and their transformations, 
the Lagrangian is constructed as seen in the following.
%

As far as the lowest-order derivative terms are concerned,
the gauge invariance and even-parity condition cast  
the Lagrangian into the following form in general;
\begin{align}
{\cal L}_{(2)}= a F_\pi^2\tr{\widehat\alpha_{\textrm{V}\mu}^2}+
b F_\pi^2\tr{\widehat\alpha_{\textrm{A}\mu}^2}+
c F_\pi^2\tr{\widehat\alpha_{\textrm{M}\mu}^2}+
d F_\pi^2\tr{(\widehat\alpha_{\textrm{A}\mu}+\widehat\alpha_{\textrm{M}\mu})^2},
\end{align}
where, 
\begin{align}
\widehat\alpha_{\textrm{V}\mu}&=\frac{1}{2}\left(\xi_\textrm{M}\widehat\alpha_{\textrm{R}\mu}\xi_\textrm{M}^\dagger+\widehat\alpha_{\textrm{L}\mu}\right), \ \
\widehat\alpha_{\textrm{A}\mu}=\frac{1}{2}\left(\xi_\textrm{M}\widehat\alpha_{\textrm{R}\mu}\xi_\textrm{M}^\dagger-\widehat\alpha_{\textrm{L}\mu}\right).
\end{align}
We can rewrite the Lagrangian by introducing $\widehat\alpha_{\pi\mu}\equiv\widehat\alpha_{\textrm{A}\mu}+\widehat\alpha_{\textrm{M}\mu}$:
\begin{align}
\mathcal{L}_{(2)}
 &  = F_\pi ^2 (a{\textrm{tr}[}\widehat\alpha _{\textrm{V}\mu}^2 ] + b{\textrm{tr}[(}\widehat\alpha _{\pi\mu}   - \widehat\alpha _{\textrm{M}\mu} {{)}}^{2} ] + c{\text{tr}[}\widehat\alpha _{\textrm{M}\mu}^2 ] + d{\text{tr}[}\widehat\alpha _{\pi\mu} ^2 ]) \nonumber\\ 
 &  = F_\pi ^2 {(}a{\text{tr}[}\widehat\alpha _{\textrm{V}\mu}^2 ] + (b + c){\text{tr}[(}\widehat\alpha _{\textrm{M}\mu}  - \frac{b}
{{b + c}}\widehat\alpha _{\pi\mu}  {{)}}^{2} ] + (d + \frac{{bc}}
{{b + c}}){\text{tr}[}\widehat\alpha _{\pi\mu} ^2 ]) .
\end{align}
By renormalizing the pion field and appropriately rewriting the coefficients we obtain
\begin{align}
\mathcal{L}_{(2)}
 &  = F_\pi ^2( {\text{tr}[}\widehat\alpha _{\pi\mu} ^2 ]+
a{\text{tr}[}\widehat\alpha _{\textrm{V}\mu}^2 ] + \beta 
 {\text{tr}[(}\widehat\alpha _{\textrm{M}\mu}  - \gamma \widehat\alpha _{\pi\mu}  )^2 ]). \label{eq:LagGHLS}
\end{align} 
The first term is nothing but the non-linear sigma model Lagrangian.
The second and third terms correspond to mass terms for vector and 
axial-vector mesons.
$\pi-A_1$ mixing is controlled by the parameter, $\gamma$,
which takes a value between $0$ and $1$.
Unless the kinetic terms of the gauge bosons are incorporated,
the Lagrangian is reduced to the non-linear sigma model
as it should be in the low-energy limit.
In fact, the equations of motion for $V_\mu$ and $A_\mu$ are given by
\begin{align}
V_\mu &=\frac{1}{2}\left(\xi_{\textrm{M}\mu}^\dagger(\alpha_{\textrm{V}\mu}-\alpha_{\textrm{M}\mu}+\gamma\alpha_{\textrm{A}\mu})\xi_{\textrm{M}\mu}+\alpha_{\textrm{V}\mu}+\alpha_{\textrm{M}\mu}-\gamma\alpha_{\textrm{A}\mu}\right),\nonumber\\
A_\mu&=\frac{1}{2}\left(\xi_{\textrm{M}\mu}^\dagger(\alpha_{\textrm{V}\mu}-\alpha_{\textrm{M}\mu}+\gamma\alpha_{\textrm{A}\mu})\xi_{\textrm{M}\mu}-\alpha_{\textrm{V}\mu}-\alpha_{\textrm{M}\mu}+\gamma\alpha_{\textrm{A}\mu}\right), \label{eq:eom}
\end{align}
where fields $\alpha_\textrm{V,A,M}$ are defined as $\widehat{\alpha}_\textrm{V,A,M}$ with 
$V_\textrm{L,R}=0$.  
Substituting Eq.(\ref{eq:eom}) into the Lagrangian Eq.(\ref{eq:LagGHLS})
leaves only the first term:
$ \mathcal{L}_{(2)}= F_\pi ^2\text{tr}[\widehat\alpha _{\pi\mu} ^2 ].$

The kinetic term of gauge boson is given by
\begin{align}
\mathcal{L}_\textrm{kin}&=-\frac{1}{{2g^2 }}{\text{tr}}[(F_{\mu \nu }^{(\textrm{L})})^2 ]-\frac{1}{{2g^2 }}{\text{tr}}[(F_{\mu \nu }^{(\textrm{R})})^2]\nonumber\\
&=-\frac{1}{{2g^2 }}{\text{tr}}[(F_{\mu \nu }^{(\textrm{V})})^2]-\frac{1}{{2g^2 }}{\text{tr}}[(F_{\mu \nu }^{(\textrm{A})})^2],
\end{align}
where
\begin{align}
 F_{\mu \nu }^{(\textrm{V})}  &= \partial _\mu  V_\nu   - \partial _\nu  V_\mu   
 - i[V_\mu,V_\nu]   - i[A_\mu, A_\nu],   \nonumber\\ 
  F_{\mu \nu }^{(\textrm{A})}  &= \partial _\mu  A_\nu   - \partial _\nu  A_\mu   
 - i[V_\mu, A_\nu]   - i[A_\mu, V_\nu],
\end{align}
with $V_\mu=(V_{\textrm{R}\mu}+V_{\textrm{L}\mu})/2$ and $A_\mu=(V_{\textrm{R}\mu}-V_{\textrm{L}\mu})/2$.

Here we parameterize the nonlinear fields as
\begin{align}
\xi_\textrm{R}(x)&=\textrm{e}^{-ip(x)}\textrm{e}^{i\sigma(x)}\textrm{e}^{i\pi(x)},\nonumber\\
\xi_\textrm{L}(x)&=\textrm{e}^{ip(x)}\textrm{e}^{i\sigma(x)}\textrm{e}^{-i\pi(x)},\nonumber\\
\xi_\textrm{M}(x)&=\textrm{e}^{ip(x)}\textrm{e}^{ip(x)}.
\end{align}
By expanding the Lagrangian up to quadratic terms of these components, 
we obtain
\begin{align}
\mathcal{L}_\text{quad} =& F_\pi ^2 \text{tr}[(a(\partial_\mu \sigma )^2  + (1 + \beta \gamma ^2 )(\partial_\mu \pi )^2  + \beta (\partial_\mu p)^2  + aV_\mu^2  + \beta A_\mu^2  \nonumber\\ 
 &  - 2\beta \gamma \partial^\mu \pi \partial_\mu p + 2\beta (\partial^\mu p - \gamma \partial^\mu \pi )A_\mu - 2a\partial^\mu \sigma V_\mu \nonumber\\
 &  - 2\beta \gamma \partial^\mu p\mathcal{A}_\mu - 2\beta \gamma A^\mu\mathcal{A}_\mu + 2(1 + \beta \gamma ^2 )\partial^\mu \pi \mathcal{A}_\mu + 2a\partial^\mu \sigma \mathcal{V}_\mu - 2aV_\mu\mathcal{V}^\mu + a\mathcal{V}_\mu^2  + (1 + \beta \gamma ^2 )\mathcal{A}_\mu^2  ],
\end{align}
with ${\cal V}_\mu=({\cal R}_\mu+{\cal L}_\mu)/2$ and 
${\cal A}_\mu=({\cal R}_\mu-{\cal L}_\mu)/2$.
Note that the field $\pi$ in this Lagrangian cannot be interpreted
as the physical pion because $\pi$ is mixed 
with $p$ through the term $- 2\beta \gamma \partial^\mu \pi \partial_\mu p$ 
and components of $\pi$ are partially absorbed by $A_\mu$.
To resolve the mixing and identify the physical pion, 
we shift and rescale the field as 
$p\rightarrow (p/\sqrt{\beta}+\gamma\pi)/F_\pi$, which leads us to
\begin{align}
 \mathcal{L}_\text{quad} =& \text{tr}[(\partial_\mu \sigma )^2  + 
 (\partial_\mu \pi )^2  + (\partial_\mu p)^2  + M_\rho\rho_\mu^2  + M_{A_1}^2 A_{1\mu}^2  + 2F_p \partial^\mu pA_\mu - 2F_\sigma\partial^\mu \sigma V_\mu \nonumber\\ 
 &  - 2F_{A_1}\partial^\mu p\mathcal{A}_\mu - 
 2g\beta\gamma F_\pi^2A_1^\mu\mathcal{A}_\mu 
+ 2F_\pi\partial^\mu \pi \mathcal{A}_\mu + 2F_\sigma\partial^\mu \sigma \mathcal{V}_\mu - 
2agF_\pi^2 V^\mu\mathcal{V}_\mu
 + F_\sigma^2\mathcal{V}_\mu^2  + (F_\pi^2 + F_{A_1}^2 )\mathcal{A}_\mu^2].
\end{align} 
Here, other fields are also rescaled as
\begin{align}
\pi&\to \frac{\pi}{F_\pi}, \hspace{1cm}\sigma\to \frac{\sigma}{\sqrt{a}F_\pi}.
\end{align}
The gauge fields, decay constants, gauge couplings and mass of the gauge fields
are expressed as
\begin{align}
\rho_\mu&=gV_\mu,\hspace{1cm}A_{1\mu}=gA_\mu,\nonumber\\
F_\sigma^2&=aF_\pi^2,
\hspace{1cm}F_p^2=\beta F_\pi^2,\hspace{1cm}
F_{A_1}^2=\beta\gamma^2F_\pi^2,\nonumber\\
M_\rho^2&=ag^2F_\pi^2,
\hspace{1cm}M_{A_1}^2=\beta g^2 F_\pi^2.
\end{align}
\setcounter{equation}{0}
\section{Wilsonian Matching for the GHLS}
\label{sec:WilsonMatch}
Since the GHLS model is an effective theory of QCD,
the parameters of the GHLS model can and should be determined from QCD.
In this section we apply the Wilsonian matching for the GHLS model \cite{Harada:2001b, Harada:2003jx}. 
Following ref.\cite{Harada:2001b, Harada:2003jx}, we adjust
the vector and axial-vector current correlators in the GHLS model 
to the correlators in QCD calculated by the operator product expansion (OPE).
Here we assume the existence of an energy scale, $\Lambda$, above $A_1$ meson mass
where both the GHLS model and the OPE of QCD are applicable.

The vector and axial-vector current correlators are defined by
\begin{align}
 i\int {{\text{d}}^4 x{\text{e}}^{ipx} } \left\langle 0 \right|TJ_\mu ^a (x)J_\nu ^b (0)\left| 0 \right\rangle  &=  - \delta ^{ab} P^{\mu \nu } \Pi _{\mathcal{V}} (Q^2 ), \nonumber\\ 
 i\int {{\text{d}}^4 x{\text{e}}^{ipx} } \left\langle 0 \right|TJ_{5\mu 
 }^a (x)J_{5\nu }^b (0)\left| 0 \right\rangle  &=  - \delta ^{ab} P^{\mu \nu } \Pi _{\mathcal{A}} (Q^2 ) ,
\end{align}
where $Q^2=-p^2$ and $P^{\mu\nu}=g^{\mu\nu}p^2-p^\mu p^\nu$.
These correlators are well described by the tree-level contributions
around the matching scale, $\Lambda$ \cite{Harada:2001b, Harada:2003jx}. 
In the GHLS model, the correlators up to ${\cal O}(p^4)$ are given by
\begin{align}
\Pi_{\mathcal{A}}^\textrm{(GHLS)}(Q^2)&=\frac{F_\pi^2(\Lambda)}{Q^2}+\frac{F_{A_1}^2(\Lambda)(1-g_{A_1}^2(\Lambda)z_4(\Lambda)/\gamma(\Lambda))}{M_{A_1}^2(\Lambda)+Q^2}-2z_2(\Lambda),\nonumber\\
\Pi_{\mathcal{V}}^\textrm{(GHLS)}(Q^2)&=\frac{F_\sigma^2(\Lambda)(1-2g^2_\rho(\Lambda)z_3(\Lambda))}{M_\rho^2(\Lambda)+Q^2}-2z_1(\Lambda),
\label{eq:crrntcrrGHLS}
\end{align}
where
$g_\rho^2(\Lambda)=g^2(\Lambda)/(1+2\kappa(\Lambda))$, $g_{A_1}^2(\Lambda)=g^2(\Lambda)/(1-2\kappa(\Lambda))$, 
$M_\rho^2(\Lambda)=g_\rho^2(\Lambda)F_\sigma^2(\Lambda)$ and $M_{A_1}^2(\Lambda)=g_{A_1}^2(\Lambda)F_p^2(\Lambda)$.
$\kappa$, $z_1$, $z_2$, $z_3$ and $z_4$ are coefficients in the 
following Lagrangian of ${\cal O}(p^4)$:
\begin{align}
{\mathcal{L}}_{(4)z}  =& \frac{\kappa }
{{2g^2 }}{\text{tr}}[F_{\mu \nu }^{(\textrm{L})} \xi _\textrm{M} F^{(\textrm{R})\mu \nu } \xi _\textrm{M}^\dag  ] + \frac{{z_1  + z_2 }}
{4}{\text{tr}}[({\mathcal{F}}_{\mu \nu }^{(\textrm{L})})^2 + ({\mathcal{F}}_{\mu \nu }^{(\textrm{R})})^2] + \frac{{z_1  - z_2 }}
{2}{\text{tr}}[{\mathcal{F}}_{\mu \nu }^{(\textrm{L})} \xi _\textrm{L}^\dag  \xi _\textrm{M} \xi _\textrm{R} {\mathcal{F}}^{(\textrm{R})\mu \nu } \xi _\textrm{R}^\dag  \xi _\textrm{M}^\dag  \xi _\textrm{L} ] \nonumber\\ 
 &  + \frac{{z_3  + z_4 }}
{4}{\text{tr}}[\xi _\textrm{L} {\mathcal{F}}_{\mu \nu }^{(\textrm{L})} \xi _\textrm{L}^\dag  F^{(\textrm{L})\mu \nu }  + \xi _\textrm{R} {\mathcal{F}}_{\mu \nu }^{(\textrm{R})} \xi _\textrm{R}^\dag  F^{(\textrm{R})\mu \nu }] + \frac{{z_3  - z_4 }}
{4}{\text{tr}}[{\mathcal{F}}_{\mu \nu }^{(\textrm{L})} \xi _\textrm{L}^\dag  \xi _\textrm{M} F^{(\textrm{R})\mu \nu } \xi _\textrm{M}^\dag  \xi _\textrm{L}  + {\mathcal{F}}_{\mu \nu }^{(\textrm{R})} \xi _\textrm{R}^\dag  \xi _\textrm{M}^\dag  F^{(\textrm{L})\mu \nu } \xi _\textrm{M} \xi _\textrm{R} ].
\end{align}
On the other hand, the OPE in QCD leads to the correlators up 
to ${\cal O}(1/Q^6)$ \cite{Shifman:1978bx,Shifman:1978by} as  
\begin{align}
\Pi_{\cal A}^\textrm{(QCD)}(Q^2)&=\frac{1}{8\pi^2}\left(
-\left(1+\frac{\alpha_s}{\pi}\right)\ln\frac{Q^2}{\mu^2}
+\frac{\pi\alpha_s}{3}\frac{\langle G_{\mu\nu}G^{\mu\nu}\rangle}{Q^4}
+\frac{1408\pi^3\alpha_s}{81}\frac{\langle\bar{q}q\rangle^2}{Q^6}\right),\nonumber\\
\Pi_{\cal V}^\textrm{(QCD)}(Q^2)&=\frac{1}{8\pi^2}\left(
-\left(1+\frac{\alpha_s}{\pi}\right)\ln\frac{Q^2}{\mu^2}
+\frac{\pi\alpha_s}{3}\frac{\langle G_{\mu\nu}G^{\mu\nu}\rangle}{Q^4}
-\frac{896\pi^3\alpha_s}{81}\frac{\langle\bar{q}q\rangle^2}{Q^6}\right),
\label{eq:crrntcrrOPE}
\end{align}
where $\mu$ is the renormalization scale of QCD.

Now let us make the Wilsonian matching of
Eq.(\ref{eq:crrntcrrGHLS}) with Eq.(\ref{eq:crrntcrrOPE}) 
at the scale $\Lambda$.
We note that the correlators in Eq.(\ref{eq:crrntcrrOPE}) explicitly depend 
on the renormalization scale $\mu$.
To eliminate the explicit $\mu$ dependence, we consider the difference, 
$\Pi_\mathcal{A}-\Pi_\mathcal{V}$, and the first derivative of 
the correlators.
Then the matching condition for the difference is
\begin{align}
\left.\Big({\Pi _{\mathcal{A}}^\textrm{(GHLS)} (Q^2 )-\Pi _{\mathcal{V}}^\textrm{(GHLS)} (Q^2 )}\Big) \right|_{Q^2  = \Lambda ^2 }  
&= \left. \left(\Pi _{\mathcal{A}}^\textrm{(QCD)} (Q^2 )-\Pi _{\mathcal{V}}^\textrm{(QCD)} (Q^2 )\right) \right|_{Q^2  = \Lambda ^2 }
=\frac{32\pi\alpha_s}{9}\frac{\langle\bar{q}q\rangle^2}{\Lambda^6}.
\end{align}
The matching conditions of the first derivative 
for each $\Pi_{\mathcal{A},\mathcal{V}}$ are given by
\begin{align}
  - Q^2 \frac{d}{{dQ^2 }}\left. {\Pi _{\mathcal{A},\mathcal{V}}^\textrm{(GHLS)} (Q^2 )} \right|_{Q^2  = \Lambda ^2 }  &= \left. { - Q^2 \frac{d}
{{dQ^2 }}\Pi _{\mathcal{A},\mathcal{V}}^\textrm{(QCD)} (Q^2 )} \right|_{Q^2  = \Lambda ^2 },
\end{align}
where
\begin{align}
\left.- Q^2 \frac{d}{{dQ^2 }}\Pi _\mathcal{A}^\textrm{(GHLS)} (Q^2 ) \right|_{Q^2  = \Lambda ^2 }
 &= \frac{{F_\pi  ^2 }(\Lambda)}
{{\Lambda^2 }} + \frac{{\Lambda^2 F_{A_1} ^2(\Lambda) (1 - 2g_{A_1}^2(\Lambda) z_4(\Lambda)/\gamma(\Lambda) )}}
{{(\Lambda^2  + M_{A_1 } ^2(\Lambda) )^2}}, \nonumber\\
- \left.Q^2 \frac{d}{{dQ^2 }}\Pi _\mathcal{V}^\textrm{(GHLS)} (Q^2 ) \right|_{Q^2  = \Lambda ^2 }
 &= \frac{{\Lambda^2 F_\sigma  ^2(\Lambda) (1 - 2g_\rho^2(\Lambda) z_3(\Lambda) )}}
{{(\Lambda^2  + M_\rho  ^2(\Lambda) )^2}},
\label{eq:match}
\end{align}
and
\begin{align}
\left.- Q^2 \frac{d}{{dQ^2 }}\Pi _\mathcal{A}^\textrm{(QCD)} (Q^2 ) \right|_{Q^2  = \Lambda ^2 }
&=\frac{1}{8\pi^2}\left(
1+\frac{\alpha_s}{\pi}
+\frac{2\pi\alpha_s}{3}\frac{\langle G_{\mu\nu}G^{\mu\nu}\rangle}{\Lambda^4}
+\frac{1408\pi^3\alpha_s}{27}\frac{\langle\bar{q}q\rangle^2}{\Lambda^6}\right),\nonumber\\
- \left.Q^2 \frac{d}{{dQ^2 }}\Pi _\mathcal{V}^\textrm{(QCD)} (Q^2 ) \right|_{Q^2  = \Lambda ^2 }
&=\frac{1}{8\pi^2}\left(1+\frac{\alpha_s}{\pi}
+\frac{2\pi\alpha_s}{3}\frac{\langle G_{\mu\nu}G^{\mu\nu}\rangle}{\Lambda^4}
-\frac{896\pi^3\alpha_s}{27}\frac{\langle\bar{q}q\rangle^2}{\Lambda^6}\right).
\label{eq:dermatch}
\end{align}

When the chiral symmetry is restored, the vector current correlator coincides 
with the axial one due to $\langle\bar{q}q\rangle\to0$:
\begin{align}
\left.\Big({\Pi _{\mathcal{A}}^\textrm{(GHLS)} (Q^2 )-\Pi _{\mathcal{V}}^\textrm{(GHLS)} (Q^2 )}\Big) \right|_{Q^2  = \Lambda ^2 }  
&\to 0.
\label{eq:restmatch}
\end{align}
The derivatives of the vector and axial-vector correlators
of Eq.(\ref{eq:dermatch}) become also identical with each other.
We emphasize that their values are finite but nonzero 
because $(\alpha_s/\pi)\langle G_{\mu\nu}G^{\mu\nu}\rangle$ may not 
vanish at the restoration point.
This implies 
\begin{align}
{F_\pi  ^2 }(\Lambda) + \frac{ F_{A_1} ^2(\Lambda) (1 - 2g_{A_1}^2(\Lambda) z_4(\Lambda)/\gamma(\Lambda) )}
{{(1  + M_{A_1 } ^2(\Lambda)/\Lambda^2 )^2}}\neq0, \nonumber\\
F_\sigma^2(\Lambda)(1-2g_\rho^2(\Lambda)z_3(\Lambda))\neq0.
\label{eq:fsig}
\end{align}
With the parameters obtained by the Wilsonian matching,
the low-energy parameters such as on-shell $\rho$ meson mass are derived by
the renormalization group.
We discuss more detail of the restoration of chiral symmetry in sec.\ref{sec:Restpattern}.
\setcounter{equation}{0}
\section{Renormalization Group Equations}
\label{sec:RGEs}
 In the HLS model, quantum effects on couplings have been studied by the
renormalization group \cite{Harada:1992bu,Harada:1999zj}.
It was pointed out that the quadratic divergence 
plays an important role in the chiral transition, especially 
in connection with the large $N_f$ QCD \cite{Harada:1999zj}. 
Quantum corrections to the bare parameters suggest a violation of 
the vector meson dominance. 
 Following this argument in the HLS model,
we consider the renormalization group for the couplings in the GHLS model to
estimate quantum effects and find fixed points. 
To this end, we introduce dimensionless parameters for convenience;
\begin{align}
x(\mu ) &= \frac{{N_f }}
{{(4\pi )^2 }}\frac{{\mu ^{2} }}{{F_\pi  ^2(\mu ) }}, 
\hspace{.5cm}G(\mu)=  \frac{{N_f }}
{{(4\pi )^2 }}g^2(\mu ). \label{dimlessparam}
\end{align}
By calculating one-loop diagrams contributing to the bare couplings
as shown in Appendix, 
we obtain renormalization group equations for the parameters as 
\begin{align}
  \mu \frac{{dx}}
{{d\mu }} = & \ x\left\{2- x\left(2 - a + 2\gamma ^2  + 2a\gamma ^2  
- \frac{{a\gamma ^2 }}{\beta } - \frac{{\beta \gamma ^2 }}
{a} - a\gamma ^4 \right) \right.\\ 
 & \ \ \ \  \left.+ \frac{3}
{2}G\left(a^2  - a\gamma ^2  - 2a^2 \gamma ^2  + \frac{{a^2 \gamma ^2 }}
{\beta } - \beta \gamma ^2  + \frac{{\beta ^2 \gamma ^2 }}
{a} + a^2 \gamma ^4 \right)\right\} \nonumber,\\
\mu \frac{{da}}
{{d\mu }} = & \ x\left(\frac{1}
{2} - 2a + \frac{{3a^2 }}
{2} + \frac{{a^2 }}
{{2\beta ^2 }} - 2a\gamma ^2  - 3a^2 \gamma ^2  + \frac{{2a^2 \gamma ^2 }}
{\beta } + \frac{{3a^2 \gamma ^4 }}
{2}\right) \nonumber\\ 
 &  + \frac{3}{2}Ga\left(1 - a^2  + \frac{a}
{\beta } + 2a\gamma ^2  + 2a^2 \gamma ^2  - \frac{{a^2 \gamma ^2 }}
{\beta } - \beta \gamma ^2  - a^2 \gamma ^4 \right) \nonumber,\\
\mu \frac{{d\beta }}{{d\mu }} = & \ x\left(2 - \frac{a}
{\beta } - 2\beta  + a\beta  - 2\beta \gamma ^2  - 2a\beta \gamma ^2  + \frac{{2\beta ^2 \gamma ^2 }}
{a} + a\beta \gamma ^4 \right) \nonumber\\ 
 &  + \frac{3}
{2}G\left( - 3a + \frac{{a^2 }}
{\beta } + 4\beta  - a^2 \beta  - a\beta \gamma ^2  + 2a^2 \beta \gamma ^2  + 2\beta ^2 \gamma ^2  - \frac{{\beta ^3 \gamma ^2 }}
{a} - a^2 \beta \gamma ^4 \right) \nonumber,\\ 
\mu \frac{{d\gamma ^2 }}
{{d\mu }} =& \ 2x\gamma ^2 \left(\frac{1}
{a} + \frac{a}{{\beta ^2 }} - \frac{2}{\beta } - \frac{a}{\beta } 
+ \frac{{a\gamma ^2 }}{\beta }\right) 
+ 3G\gamma ^2 \left( - 1 - a - \frac{{a^2 }}{{\beta ^2 }} + \frac{{2a}}
{\beta } + \frac{{a^2 }}{\beta } + a\gamma ^2  - \frac{{a^2 \gamma ^2 }}
{\beta }\right)\nonumber,\\
\mu \frac{{dG}}
{{d\mu }} =& - G^2 \left\{\frac{{44}}{3} - \frac{1}
{{24}}\left(5 + a^2  + \frac{{a^2 }}
{{\beta ^2 }} - \frac{{2a}}
{\beta } - 2a\gamma ^2  - 2a^2 \gamma ^2  + \frac{{2a^2 \gamma ^2 }}
{\beta } - 2\beta \gamma ^2  + \frac{{2\beta ^2 \gamma ^2 }}
{a} + a^2 \gamma ^4 \right)\right\}.
\label{RGEs}
\end{align}
If we take a limit of $\beta\to \infty$ and $\gamma \to 0$,
the above RGEs for $x$ and $a$ do not receive contributions from the $A_1$ meson
and are reduced to those in the HLS model.
However, it is not the case for $G$, because the number of the gauge fields in the GHLS model
is different from that in the HLS model and 
the loop contributions of the field $p$ remain in the limit.
This is due to the cancellation between
the overall decay constants in the Lagrangian and 
denominators in the power series of the fields.\
In the next section we examine these equations for fixed points to discuss the properties
on the chiral-transition point.
\setcounter{equation}{0}
\section{Restoration Pattern}
\label{sec:Restpattern}
The chiral restoration is characterized by the condition that
the on-shell pion decay constant vanishes, $F_\pi(0)\to0$. 
That is to say the non-vanishing of
the dimensionless parameter $x(\mu)$ defined in Eq.(\ref{dimlessparam})
in the low-energy limit:
\begin{equation}
x(0) \neq 0. \label{phasetrans}
\end{equation}
Inversely, $x(0)=0$ means a broken phase because of $F_\pi(0)\neq0$
and is excluded from the parameter sets associated with the restoration point.
In addition to Eq.(\ref{phasetrans}),
we analyze the restoration patterns of chiral symmetry using the vector 
and axial-vector current correlators.
When the chiral symmetry is restored,
the vector current correlator should be equal to the axial vector one,
not only at the matching scale $\Lambda$ as shown in Eq.(\ref{eq:restmatch}) but also
at the low-energy region less than $\Lambda$.
Accordingly, at the restoration point
we require equality of these correlators;
\begin{align}
\Pi_{\mathcal{V}}(Q^2)=\Pi_{\mathcal{A}}(Q^2),
\label{eq.cor}
\end{align}
for $Q^2\le \Lambda^2$.
Since the HLS model includes of only $\pi$ and $\rho$,  
the equality of correlators and Wilsonian matching uniquely bring about
the VM scenario \cite{Harada:2001, Harada:2003jx}.
In the GHLS model, however, other possibilities exist. 
Actually considering the restraints on the decay constants 
of Eq.(\ref{eq:fsig}), Eq.(\ref{eq.cor}) gives $z_1(\mu)=z_2(\mu)$ and 
preserves the following three possibilities:
\begin{align}
({\rm I})\ \ &M_\rho^2(\mu)=M_{A_1}^2(\mu), \ \ 
 F_{A_1}^2(\mu)(1-2g^2_{A_1}(\mu)z_4(\mu))=F_{\sigma}^2(\mu)(1-2g^2_\rho(\mu)z_3(\mu)) 
\ \text{ and }\ F_\pi^2(\mu)=0, \label{stcondition}\\
({\rm II})\ \ &M_\rho^2(\mu)=0,\ \ 
 F_\sigma^2(\mu)(1-2g^2_\rho(\mu)z_3(\mu))=F_\pi^2(\mu) 
\ \text{ and }\
      F_{A_1}^2(\mu)(1-2g^2_{A_1}(\mu)z_4(\mu)/\gamma(\mu))=0, \label{VMcondition}\\
({\rm III})\ \ &M_\rho(\mu)=M_{A_1}(\mu)=0 \ \text{ and }\
  F_\pi^2(\mu)=F_\sigma^2(\mu)(1-2g^2_\rho(\mu) z_3(\mu))-F_{A_1}^2(\mu)(1-2g_{A_1}^2(\mu)z_4(\mu)/\gamma(\mu)). 
\label{Intermediatecondition}
\end{align}
In order to determine whether each of the above possibilities can be realized,
we require that the restoration condition is satisfied for any energy below the matching scale.
This demands that each of Eqs.~(\ref{stcondition})-(\ref{Intermediatecondition}) is
renormalization-group invariant.

Let us begin with the case (I).
Eq.~(\ref{stcondition}), $F_\pi^2(\mu)=0$ implies that
the power expansion of $p^2/(4\pi F_\pi(\Lambda))^2$ breaks down.
Therefore, the first possibility is inconsistent with the one-loop approximation
adopted in this analysis and is not studied further.

Next, we consider the case (II).
From Eq.~(\ref{VMcondition}) we obtain
\begin{align}
g^2=0, ~~~~a=1,~~~~\beta\gamma^2=0, \label{VMcondition2}
\end{align}
for any $\mu$ less than the matching scale. 
Since the renormalization group equations for these quantities 
are rendered as
\begin{align}
\mu \frac{{dG}}{{d\mu }} = 0,\ \ 
\mu \frac{{da}}{{d\mu }} = x\frac{1}{{2\beta ^2 }},\ \ 
\mu \frac{{d(\beta\gamma ^2) }}{{d\mu }} =   x\gamma ^2(-4 +\frac{1}{\beta}+ 2\gamma^2),\ \ 
\mu \frac{d(z_1-z_2)}{d\mu}=\frac{{N_f }}
{{(4\pi )^2 }}\frac{1}{24\beta^2},
\end{align}
Eq.(\ref{VMcondition2}) with $z_1(\mu)=z_2(\mu)$ 
results in $\beta\to\infty$ and $\gamma = 0$ for any $\mu$.
As we take a low energy limit $\mu\to0$, other parameters are also 
expected to be in a fixed point. Considering the above conditions together
with the RGE for $x$
\begin{equation}
\mu \frac{{dx}}{{d\mu }} = x(2 - x),\ \ 
\end{equation}
the non-trivial fixed point is found to be
$(x,a,\beta,\gamma^2,G)=(2, 1, \infty, 0, 0)$.
This fixed point corresponds to the VM, in which $M_\rho\to M_\pi=0$ and $F_\sigma/F_\pi\to1$.
We note that a combination of $\beta\gamma^2$ remains to be zero 
as $\beta\to\infty$
and $\gamma\to0$ from Eq.(\ref{VMcondition2}) and hence $F_{A_1} = 0$.
We can say that $A_1$ is decoupled to the system in this case and
the GHLS model is reduced to the HLS model at the restoration point.

Finally, we discuss the  case (III).
Note that the second condition of Eq.(\ref{Intermediatecondition}) is nothing but 
the first Weinberg sum rule which is saturated by the pole terms \cite{Weinberg:1969hw}.
From Eq.(\ref{Intermediatecondition}) we obtain 
\begin{align}
g^2=0 \ \text{  and   }\ a-\beta\gamma^2-1=0, \label{cond2}
\end{align}
for any $\mu$ less than the matching scale.
The  RGEs for these quantities are given by
\begin{align}
\mu \frac{{d G}}{{d\mu }} = 0,\ \ 
\mu \frac{{d(a - \beta \gamma ^2  - 1)}}{{d\mu }} &= 
 - x\frac{{(1 - \gamma ^2 )a^2  - 2a + 1}}{{2(1 - a)^2 }}, \ \ 
\mu \frac{{d(z_1  - z_2 )}}{{d\mu }} = \frac{{N_f }}
{{(4\pi )^2 }}\frac{{(1 - \gamma ^2 )a^2  - 2a + 1}}{{24(1 - a)^2 }}.
\end{align}
From these, we can see that
a parameter set of $a=1/(1-\gamma)$ and $\beta=1/((1-\gamma)\gamma)$
makes Eq.(\ref{cond2}) invariant under the renormalization group. 
Here we note that this parameter set
satisfies the first and second Weinberg sum rules at ${\cal O}(p^2)$ 
even if we take a nonzero $g$ \cite{Harada:2005br}.
Under these conditions, remaining RGEs for $x$ and $\gamma^2$ become
\begin{align}
\mu \frac{{d x}}{{d\mu }} &= x(2 - (1 - 2\gamma  + 3\gamma ^2 )x), \ \ 
\mu \frac{{d\gamma ^2 }}{{d\mu }} = 2\gamma ^2 (1 - \gamma )(1 - 3\gamma )x.
\label{intermediatefixed}
\end{align}
The parameters $x$ and $\gamma$ are
expected to be in a fixed point in the low energy limit of $\mu\to0$.
Except for the trivial fixed point of $x=0$ with an arbitrary $\gamma$, 
we find three fixed points from Eq.(\ref{intermediatefixed}), 
\begin{align}
(x,\gamma)&=
\begin{cases}
(1, 1)  & \text{~~~~~(the standard type)}\\
(2, 0)  & \text{~~~~~(the VM type)}\\
(3, 1/3)& \text{~~ (the intermediate type)}\label{IMfix}.
\end{cases}
\end{align}
These fixed points are classified into the restoration types noted above
according to the value of $\gamma$, because the $\pi$-$A_1$ mixing are 
basically controlled by $\gamma$ as discussed in the next subsection.
The parameters below the matching scale and the fixed points are 
summarized for each restoration pattern in Table \ref{tab:fp}.

Under the standard scenario, if $a$ was in a fixed point at the matching scale,
$F_\sigma(\Lambda)$ would diverge 
and be inconsistent with Eq.(\ref{eq:fsig}) because 
$F_\sigma(\Lambda)$ must be finite.
Thus we do not claim that the parameters are fixed on a fixed point at the 
matching scale.
Under the VM scenario in the case of (II), the parameters are on a fixed point 
and $\gamma$ keeps taking zero for  $\mu$ lower than the matching scale.
However, for the VM in the case of (III), $\gamma$ does
not vanish for a non-zero $\mu$ and is not on any fixed points.
Consequently, in the case of (II) $A_1$ is decoupled in the bare Lagrangian,
while for the VM in case of (III) $A_1$ remains to be 
dynamical at least near the matching scale. 
This difference is expected to reflect on the spectrum of 
$A_1$ as the chiral symmetry is restored.

Although the parameters may not be located on the fixed point at the matching scale,
the couplings must be on the critical line along which the 
renormalization-group flow enters into an infrared fixed point satisfying $x(0)\neq0$.
The intermediate fixed point is on the critical line but 
an UV fixed point. It requires a fine adjustment to keep the point
realized in IR region, but no reason exists to justify such an adjustment.
The restoration conditions put severe constraints on the couplings,
and without these conditions all of the fixed points become unstable. 
Unless Eq.(\ref{phasetrans}) is satisfied,
a fixed point of the vector realization defined as
$x=0$ with Eq.(\ref{eq.cor}) will be achieved \cite{Georgi:1989gp}.

It turns out that
the gauge coupling becomes zero, $g=0$, at the all fixed points 
satisfying the restoration condition. This fact implies that $\rho$ and $A_1$ 
get lighter as the chiral symmetry is restored and eventually become massless 
at the restoration point. 
This result is consistent with the Brown-Rho scaling \cite{Brown:1991kk}.

\begin{table}
\begin{center}
\begin{tabular}{|l|c|c|c|c|c|c|c|c|}\hline
             & \raisebox{-.6em}[0pt][0pt]{$0<\mu\le\Lambda$}
 &\multicolumn{6}{|c|}{fixed point $( \mu\to0 )$}\\ \cline{3-8}
             & 
             & $x$ & $a$ & $\beta$ & $\gamma$ &  $\psi$&  $g_{\gamma\pi\pi}$\\ \hline
(II) VM      & $a=1$, $\beta=\infty$, $\gamma=0$ ( $\beta\gamma^2=0$ )
             & $2$ & $1$        & $\infty$ & $0$ & $0$     &$1/2$ \\ \hline
(III) VM     &   
             &$2$ & $1$        & $\infty$ & $0$ & $0$     &$1/2$ \\ 
Standard     & \raisebox{0em}[0pt][0pt]
{$  \begin{cases}a=1/(1-\gamma) ,\\
\beta=1/(\gamma(1-\gamma))      \end{cases}$}
             &~$1$~ & $\infty$   & $\infty$ & $1$ & $\pi/2$ & $0$ \\ 
Intermediate &
             &$3$ & $3/2$      & $9/2$    & $1/3$ & $\arctan\sqrt{2}$   &$1/3$\\ \hline
\end{tabular}
\end{center}
\caption{Parameters and their fixed points for the restoration patterns.
The gauge coupling is zero, $g=0$, for these restoration patterns.
In the intermediate scenario, the fixed point is not realized in the 
 limit of $\mu\to0$ unless the parameters are adjusted at the matching scale.
}
\label{tab:fp}
\end{table}

\subsection{The mixing angle}

When mesons are classified by the light-front charge of the chiral group $SU(N_f) _L\times SU(N_f)_R$ in the broken phase, 
the spontaneous breaking of the chiral symmetry mixes the pseudo-scalar meson
with the longitudinal part of the axial-vector meson while the scalar and vector mesons remain to be pure representations:

\begin{eqnarray}
  &&|\sigma\rangle = |(N_f^2-1,1) \oplus (1,N_f^2-1)\rangle, \nonumber\\
  &&|\pi\rangle = |(N_f,N_f^*) \oplus (N_f^*,N_f)\rangle \sin\psi + |(N_f^2-1,1) \oplus (1,8)\rangle \cos\psi, \nonumber\\
  &&|\rho\rangle = |(N_f,N_f^*) \oplus (N_f^*,N_f)\rangle, \nonumber\\
  &&|A_1\rangle = |(N_f,N_f^*) \oplus (N_f^*,N_f)\rangle \cos\psi - |(N_f^2-1,1) \oplus (1,N_f^2-1)\rangle \sin\psi.
\end{eqnarray}
The experimental values of the decay constants suggest that the
mixing angle, $\psi$, takes 
$\psi\simeq \pi/4$ assuming two light flavors \cite{Gilman:1960,Weinberg:1969}.
As the chiral symmetry is restored, the mixing is expected to disappear
and physical states become pure representations
because the light-front chiral charge commutes with the Hamiltonian \cite{Yamawaki:1998cy}.
It is non-trivial to which representation the pion belongs.
In the case that the pion belongs to $(N_f,N_f^*)\oplus(N_f^*,N_f)$ with 
$\psi=\pi/2$, the scalar meson becomes the chiral partner of
the pion and we call it the standard scenario of the chiral restoration. 
In the case that the pion belongs to $(N_f^2-1,1)\oplus(1,N_f^2-1)$ with
$\psi=0$, the pion and the longitudinal mode of the $\rho$ meson
form a doublet, which is called the VM scenario \cite{Harada:2001}.

In this section, we determine the mixing angle at the restoration point for each restoration pattern classified in the above.
We define the mixing angle in the GHLS model 
as the ratio of the pole residues for $\pi$ and $A_1$ in the axial-vector current correlator.
These residues 
correspond to $\pi$ and $A_1$ decay constants squared and the mixing angle
is expressed as
\begin{align}
\tan^2\psi=\frac{F_{A_1}^2(M_{A_1}^2)(1-g_{A_1}^2(M_{A_1}^2)z_4(M_{A_1}^2)/\gamma(M_{A_1}^2))}{F_\pi^2(0)}=\beta(M_{A_1}^2)\gamma^2(M_{A_1}^2)(1-g_{A_1}^2(M_{A_1}^2)z_4(M_{A_1}^2)/\gamma(M_{A_1}^2)).
\end{align}
At the restoration point, this expression becomes
\begin{align}
\tan^2\psi=\beta(0)\gamma^2(0),
\end{align}
due to $M_{A_1}=0$ and $g_{A_1}=0$.
Therefore the mixing angle is controlled by the infrared behavior
of the parameters near the restoration point.
Corresponding to the three fixed points discussed in the previous section, 
the mixing angle at the restoration point is given by
\begin{align}
\text{Standard}:&\tan\psi\to\infty, \ \ 
\text{VM}:\tan\psi\to0, \ \ 
\text{Intermediate}:\tan\psi\to\sqrt{2}.
\end{align}
As is expected, the mixing disappears at the standard and VM fixed points, 
but survives at the intermediate fixed point.
The intermediate fixed point should be excluded if we require that the 
mixing of $\pi$ and $A_1$ is resolved at the restoration point.

\subsection{Pion form factor and vector meson dominance}
The vector meson dominance (VMD) for the pion form factor is well established 
in the vacuum \cite{Sakurai:1969}.
The pion form factor in the GHLS model is written 
 up to ${\cal O}(p^2)$ as
\begin{align}
F_{\text{V}}^{\pi ^ \pm  } (p^2 ) = \left(1 - \frac{a}
{2}(1 - \gamma ^2 )\right) 
+ \frac{a}{2}(1 - \gamma ^2 )\frac{{M_\rho  ^2 }}{{M_\rho  ^2  - p^2 }}.
\end{align}
The first term is the direct term, $g_{\gamma\pi\pi}$, and 
the second term is the $\rho$ meson exchange term.
The VMD implies
\begin{align}
g_{\gamma\pi\pi}=1 - \frac{a}{2}(1 - \gamma ^2 ) \approx 0.
\end{align}
At the chiral restoration point, this coupling is
expressed as 
\[g_{\gamma\pi\pi}(\mu)=\frac{1-\gamma(\mu)}{2}\] 
and its behavior at 
low energy is different according to the restoration patterns classified above. 
Under the VM scenario in the case of (II) $g_{\gamma\pi\pi}$ takes 
a constant value of 1/2. In the case of (III), the coupling runs depending
on the parameter  $\gamma$ and 
shows (in)violation of the VMD in the infrared limit as
\begin{align}
g_{\gamma\pi\pi} &\to
\begin{cases}
0  & \text{~~~~~(the standard type)}\\
1/2  & \text{~~~~~(the VM type)}\\
1/3 & \text{~~~~~(the intermediate type)}
\end{cases}
\ \ \ {\rm as} \ \mu\to0. 
\end{align}
In the standard scenario the VMD is satisfied.
The violation of VMD appears in the VM and Intermediate fixed points.

On the other hand, $g_{\rho\pi\pi}$ coupling is written by
\begin{align}
g_{\rho\pi\pi}=\frac{g}{2}a(1-\gamma^2).
\end{align}
The coupling $g_{\rho\pi\pi}$ becomes weaker as chiral symmetry 
is restored for all restoration patterns
because the gauge coupling diminishes.
This implies that $\rho$ becomes stable for the decay into two pions
near the restoration point
and makes a sharp peak in the dilepton spectrum on the background from pions,
universally. 

The difference among the restoration patterns
may be reflected on the pion background through the violation of the VMD.
The pion background in the dilepton spectrum
is suppressed in the standard scenario, while
the large violation of the VMD leads to an enhancement of the
background in the VM scenario.
Thus observations of the dilepton spectrum in the vector channel
near the restoration point
will ascertain the validity of the GHLS model adopted here 
as well as the true restoration pattern in the real world.
\setcounter{equation}{0}
\section{Summary}
\label{sec:summary}
In this paper we have studied possible patterns of chiral restoration employing the GHLS model, which includes the axial-vector meson in addition to the pseudo-scalar and vector mesons. 
First, we obtained constraints on the parameters of the GHLS by matching the vector and axial-vector correlators of the GHLS model with those in the operator product expansion of QCD at a matching scale, $\Lambda$.
Next, we derived renormalization group equations for the GHLS model by calculating one-loop corrections to the two-point functions.
Then, we examined behavior of the parameters at low energy and
searched for fixed points of them at the restoration point by
the obtained renormalization group equations.
As the restoration conditions, we required the equality of
the vector and axial-vector correlators and the vanishing of
the on-shell pion decay constant.

We found four patterns of the running parameters at the restoration point
which flow to three fixed points.
The first fixed point correspond to the \lq standard scenario\rq, in which the vector and axial-vector mesons are degenerate. 
The second one is the \lq vector manifestation scenario\rq, in which the longitudinal part of the vector meson and the pseudo-scalar meson are degenerate.
In both these scenarios, 
the mixing of two representations associated with $\pi$ and $A_1$ mesons
diminishes at the restoration point as expected.
On the other hand, the third fixed-point is of a new type, 
which can be regarded as the \lq intermediate scenario\rq\ in a 
sense that the mixing remains even at the restoration point.
It is noted, however, that this fixed point is not an infrared one
and will not be achieved unless the parameters are perfectly adjusted 
to the point at the matching scale. 

In each of the restoration patterns,
both $\rho$ and $A_1$ are expected to become massless, since the gauge coupling vanishes at the restoration point. 
It is further expected that the vector meson becomes stable because the coupling constant, $g_{\rho\pi\pi}$, also vanishes and the decay into pions is suppressed
at the restoration point. 
Here we emphasize that these results are universal to all restoration patterns.
These three scenarios, however, are different in the violation of the vector meson dominance.
This difference will be reflected on the pion background in the spectrum of dilepton decay from the vector meson.
Two restoration patterns running into the VM fixed point are
also distinguished through the $A_1$ spectrum near the restoration point
owing to the different behavior of parameters at low energy.

Now, a crucial question is whether these fixed points are physically realized or not.
It is determined if one can reach the restoration points continuously as one approaches the critical temperature or density of chiral restoration.
The application to the system of finite temperature and/or density is our future problem.

We would like to mention here possible phenomenological consequences of the results of the present work.
The dropping of the $\rho$ meson mass implies the threshold enhancement in the $\rho$ meson spectrum at some temperature and/or density because the $\rho$ meson mass is expected to coincide with twice the pion mass,
which is analogous to the enhancement of the sigma spectrum \cite{Hatsuda:1985eb}
in the standard scenario studied within the linear sigma model 
\cite{Chiku:1998kd, Hidaka:2003mm}.
(Here, explicit breaking of chiral symmetry is taken into consideration and the pion is supposed to have a small finite mass.) 

Another point to be mentioned is that the GHLS model does not include the scalar meson, which becomes the chiral partner of the pseudo-scalar meson in the standard scenario.
To include the scalar meson in the analysis is therefore another important future problem.
\begin{center}
{\bf \large Acknowledgment}
\end{center}
We would like to thank Prof. M. Harada for useful discussions.
\begin{center}
{\bf Note added}
\end{center}
After finishing up this work we became aware of similarly work by Harada and Sasaki \cite{Harada:2005br}.
Although their analysis takes a different parameterization and gauge fixing from ours,
obtained renormalization group equations and results are consistent with 
our work.

\appendix
\renewcommand{\theequation}{\Alph{section}.\arabic{equation}}
\section{Gauge fixing}
We take a gauge fixing to calculate quantum corrections.
We perform a $R_\xi$ like BRS quantization.
The BRS transformations are defined by
\begin{align}
 {\bm{\delta }}_\textrm{B} \xi _\textrm{V,A}  &= iC_\textrm{V} \xi _\textrm{V,A}  + iC_\textrm{A} \xi _\textrm{A,V} , \nonumber\\ 
 {\bm{\delta }}_\textrm{B} \xi _\textrm{M}  &= i(C_\textrm{V}  - C_\textrm{A} )\xi _\textrm{M}  - i\xi _\textrm{M} (C_\textrm{V}  + C_\textrm{A} ), \nonumber\\ 
 {\bm{\delta }}_\textrm{B} V_\mu  &= \partial_\mu C_\textrm{V}  + i[C_\textrm{V} ,V_\mu ] + i[C_\textrm{A} 
 ,A_\mu ] = D_\mu C_\textrm{V} , \nonumber\\ 
 {\bm{\delta }}_\textrm{B} A_\mu  &= \partial_\mu C_\textrm{A}  + i[C_\textrm{V} ,A_\mu ] + i[C_\textrm{A} 
 ,V_\mu ] = D_\mu C_\textrm{A} , \nonumber\\ 
 {\bm{\delta }}_\textrm{B} C_\textrm{V,A}  &= i(C_\textrm{V} C_\textrm{V,A}  + C_\textrm{A} C_\textrm{A,V} ), \nonumber\\ 
 {\bm{\delta }}_\textrm{B} \bar C_\textrm{V,A}  &= iB_\textrm{V,A} , \nonumber\\ 
 {\bm{\delta }}_\textrm{B} B_\textrm{V,A}  &= 0, 
\end{align}
where $C_\textrm{V,A}$ and $\bar C_\textrm{V,A}$ are the Faddeev Popov ghost field, 
$B_\textrm{V,A}$ is the Nakanishi-Lautrup (NL) field, and
\begin{align}
 \xi _\textrm{V,A} & = \frac{1}
{2}(\xi _\textrm{R}  \pm \xi _\textrm{L} ). 
\end{align}
We choose the gauge fixing function as follows
\begin{align}
 & F = {\text{tr}}[\bar C_\textrm{V} (\partial^\mu V_\mu  + \alpha _\textrm{V} G_\textrm{V}  + \frac{1}
{2}\alpha _\textrm{V} B_\textrm{V} )]+\text{tr}[\bar C_\textrm{A} (\partial^\mu A_\mu  + \alpha _\textrm{A} G_\textrm{A}  + \frac{1}
{2}\alpha _\textrm{A} B_\textrm{A} )],
\end{align} 
where
\begin{align}
 G_\textrm{V} & = \frac{{aF_\pi  ^2 }}{{2i}}(\xi _\textrm{V}  - \xi _\textrm{V}^{_\dag  } ), \nonumber\\ 
 G_\textrm{A}  &= \frac{{\beta F_\pi  ^2 }}
{{2i}}(\gamma (\xi _\textrm{A}  - \xi _\textrm{A}^{_\dag  }  + \xi _{\textrm{M} - } ) - \xi _{\textrm{M} - } ),\nonumber\\
\xi _{\textrm{M} \pm }  &= \frac{1}{2}(\xi _\textrm{M}  \pm \xi _\textrm{M}^{_\dag  } ).
\end{align}
We obtain the gauge-fixing and FP terms:
\begin{align}
&\mathcal{L}_{{\text{GF}}}  + \mathcal{L}_{{\text{FP}}}  =  - i{\bm{\delta }}_\textrm{B} F,\nonumber\\
  \mathcal{L}_{{\text{GF}}}  &= 2{\text{tr}}[B_\textrm{V} (\partial^\mu V_\mu  + \alpha _\textrm{V} G_\textrm{V}  + \frac{1}
{2}\alpha _\textrm{V} B_\textrm{V} )] + 2{\text{tr}}[B_\textrm{A} (\partial^\mu A_\mu  + \alpha _\textrm{A} G_\textrm{A}  + \frac{1}
{2}\alpha _\textrm{A} B_\textrm{A} )], \nonumber\\ 
 \mathcal{L}_{{\text{FP}}}  &= 2i{\text{tr}}[\bar C_\textrm{V} (\partial^\mu 
 D_\mu C_\textrm{V}  + \alpha _\textrm{V} G'_\textrm{V} )] + 2i{\text{tr}}\bar C_\textrm{A} (\partial^\mu 
 D_\mu C_\textrm{A}  + \alpha _\textrm{A} G'_\textrm{A} )], \nonumber\\ 
 G_\textrm{V} ' &= {\bm{\delta }}_\textrm{B} G_\textrm{V}  = \frac{{ aF_\pi  ^2 }}
{2}(C_\textrm{V} \xi _\textrm{V}  + \xi _\textrm{V}^{_\dag  } C_\textrm{V}  + C_\textrm{A} \xi _\textrm{A}  + \xi _\textrm{A}^{_\dag  } C_\textrm{A} ), \nonumber\\ 
 G_\textrm{A} ' &= {\bm{\delta }}_\textrm{B} G_\textrm{A}  = \frac{{ \beta F_\pi  ^2 }}
{2}(\gamma (C_\textrm{V} \xi _\textrm{A}  + \xi _\textrm{A}^{_\dag  } C_\textrm{V}  + C_\textrm{A} \xi _\textrm{V}  + \xi _\textrm{V}^{_\dag  } C_\textrm{A}  + C_\textrm{V} \xi _{\textrm{M} - }  - \xi _{\textrm{M} - } C_\textrm{V}  - C_\textrm{A} \xi _{\textrm{M} + }  - \xi _{\textrm{M} + } C_\textrm{A} ) \nonumber\\ 
&~~~~~~~~~~~~~~~~~~~~~~  - (C_\textrm{V} \xi _{\textrm{M} - }  - \xi _{\textrm{M} - } C_\textrm{V}  - C_\textrm{A} \xi _{\textrm{M} + }  - \xi _{\textrm{M} + } C_\textrm{A} )).
\end{align} 
By integrating $B$, $\mathcal{L}_{{\text{GF}}}$ becomes
\begin{align}
  \mathcal{L}_{{\text{GF}}} & =  - \frac{1}
{{\alpha _\textrm{V} }}{\text{tr}}[(\partial^\mu V_\mu  + \alpha _\textrm{V} (G_\textrm{V}  - \frac{1}
{N_f}{\text{tr}}[G_\textrm{V} ])^2 ] - \frac{1}
{{\alpha _\textrm{A} }}{\text{tr}}[(\partial^\mu A_\mu  + \alpha _\textrm{A} (G_\textrm{A}  - \frac{1}
{N_f}{\text{tr}}[G_\textrm{A} ]))^2 ] \nonumber\\ 
 &  =  - \frac{1}
{{\alpha _\textrm{V} }}{\text{tr}}[(\partial^\mu V_\mu )^{\text{2}}] - \frac{1}
{{\alpha _\textrm{A} }}{\text{tr}}[(\partial^\mu A_\mu )^2 ] - 2{\text{tr}}[\partial^\mu V_\mu G_\textrm{V} ] - 2{\text{tr}}[\partial^\mu A_\mu G_\textrm{A} ] \nonumber\\ 
 &~~~  - \alpha _\textrm{V} ({\text{tr}}[G_\textrm{V} ^2 ] - \frac{1}
{N_f}{\text{tr}}[G_\textrm{V} ]^2 ) - \alpha _\textrm{A} ({\text{tr}}[G_\textrm{A} ^2 ] - \frac{1}{N_f}{\text{tr}}[G_\textrm{A} ]^2 ),
\end{align} 
where $N_f$ is the number of flavor.
\section{Renormalization}
Counter terms from ${\cal O}(p^4)$ Lagrangian are also necessary to cancel the 
divergences of one-loop diagrams in addition to counter terms from ${\cal 
O}(p^2)$ Lagrangian.
First, we define the renormalized parameters in the ${\cal O}(p^2)$ Lagrangian denoted with the subscript r as:
\begin{align}
\pi&=Z_\pi^{1/2}\pi_\textrm{r},~~~~~ \sigma=Z_\sigma^{1/2}\sigma_\textrm{r},~~~~~
 p=Z_p^{1/2}p_\textrm{r}+\delta_{\pi p}\pi_\textrm{r}, \nonumber\\
\rho&=Z_\rho^{1/2}\rho_\textrm{r},~~~~~ A_1=Z_{A_1}^{1/2}A_{1\textrm{r}},~~~ g^2=Z_gg^2_\textrm{r} \nonumber\\
F_\pi^2&=F_{\pi \textrm{r}}^2+\delta_{F_\pi^2},~~~ F_\sigma^2=F_{\sigma \textrm{r}}^2+\delta_{F_\sigma^2},~~~
F_p^2=F_{p \textrm{r}}^2+\delta_{F_p^2},~~~F_{A_1}^2=F_{A_1 \textrm{r}}^2+\delta_{F_{A_1}^2},\nonumber\\
\end{align}
where $\delta_{\pi p}$ is necessary to resolve $\pi-p$ mixing in the 
loop level. The counter term is defined $\delta_\phi=1-Z_\phi$, where 
$\phi=\pi$, $p$, $\rho$, $A_1$, $g$.
$Z_\rho=Z_{A_1}$ or $\delta_{Z_\rho}=\delta_{Z_{A_1}}$ is satisfied by the parity invariance of the Lagrangian.
We explicitly evaluate only the counter terms of ${\cal O}(p^4)$ related to the vector and axial-vector  current correlators. The relevant Lagrangian is given by
\begin{align}
{\mathcal{L}}_{(4)z}  =& \frac{\kappa }
{{2g^2 }}{\text{tr}}[F_{\mu \nu }^{(\textrm{L})} \xi _\textrm{M} F^{(\textrm{R})\mu \nu } \xi _\textrm{M}^\dag  ] + \frac{{z_1  + z_2 }}
{4}{\text{tr}}[({\mathcal{F}}_{\mu \nu }^{(\textrm{L})})^2 + ({\mathcal{F}}_{\mu \nu }^{(\textrm{R})})^2] + \frac{{z_1  - z_2 }}
{2}{\text{tr}}[{\mathcal{F}}_{\mu \nu }^{(\textrm{L})} \xi _\textrm{L}^\dag  \xi _\textrm{M} \xi _\textrm{R} {\mathcal{F}}^{(\textrm{R})\mu \nu } \xi _\textrm{R}^\dag  \xi _\textrm{M}^\dag  \xi _\textrm{L} ] \nonumber\\ 
 &  + \frac{{z_3  + z_4 }}
{4}{\text{tr}}[\xi _\textrm{L} {\mathcal{F}}_{\mu \nu }^{(\textrm{L})} \xi _\textrm{L}^\dag  F^{(\textrm{L})\mu \nu }  + \xi _\textrm{R} {\mathcal{F}}_{\mu \nu }^{(\textrm{R})} \xi _\textrm{R}^\dag  F^{(\textrm{R})\mu \nu }] + \frac{{z_3  - z_4 }}
{4}{\text{tr}}[{\mathcal{F}}_{\mu \nu }^{(\textrm{L})} \xi _\textrm{L}^\dag  \xi _\textrm{M} F^{(\textrm{R})\mu \nu } \xi _\textrm{M}^\dag  \xi _\textrm{L}  + {\mathcal{F}}_{\mu \nu }^{(\textrm{R})} \xi _\textrm{R}^\dag  \xi _\textrm{M}^\dag  F^{(\textrm{L})\mu \nu } \xi _\textrm{M} \xi _\textrm{R} ].
\end{align}
We define renormalized parameters
\begin{align}
\kappa Z_\rho=\kappa_\textrm{r}+\delta_\kappa, ~~~~~~~ z_{1,2}=z_{\textrm{r} 1,2}+\delta_{z_{1,2}},~~~~~~~
z_{3,4}Z_\rho^\frac{1}{2}Z_g^\frac{1}{2}=z_{\textrm{r} 3,4}+\delta_{z_{3,4}}.
\end{align}
\section{One-loop corrections}
In this appendix, we evaluate some two point functions and their divergent part.
We define $A(m^2)$, $B(p,m_1^2,m_2^2)$, $B^{\mu}(p,m_1^2,m_2^2)$, 
$B^{\mu\nu}(p,m_1^2, m_2^2)$, $\tilde B^{\mu\nu}(p,m_1^2, m_2^2)$ 
as follows:
\begin{align}
 A(m^2 ) &= {\int\frac{\textrm{d}^4k}{i(2\pi)^4}}\frac{1}
{{m^2  - k^2 }}, \nonumber\\ 
 B(p,m_1 ^2 ,m_2 ^2 ) &= {\int\frac{\textrm{d}^4k}{i(2\pi)^4}}\frac{1}
{{(m_1 ^2  - (k - p)^2 )(m_2 ^2  - k^2 )}}, \nonumber\\ 
 B^\mu  (p,m_1 ^2 ,m_2 ^2 ) &= {\int\frac{\textrm{d}^4k}{i(2\pi)^4}}\frac{{k^\mu  }}
{{(m_1 ^2  - (k - p)^2 )(m_2 ^2  - k^2 )}}, \nonumber\\ 
 \tilde B^{\mu \nu } (p,m_1 ^2 ,m_2 ^2 ) &= {\int\frac{\textrm{d}^4k}{i(2\pi)^4}}\frac{{k^\mu  k^\nu  }}
 {{(m_1 ^2  - (k - p)^2 )(m_2 ^2  - k^2 )}}, \nonumber\\ 
 B^{\mu \nu } (p,m_1 ^2 ,m_2 ^2 ) &= {\int\frac{\textrm{d}^4k}{i(2\pi)^4}}\frac{{(2k - p)^\mu  (2k - p)^\nu  }}
{{(m_1 ^2  - (k - p)^2 )(m_2 ^2  - k^2 )}}.
\end{align}
The quadratic divergence plays an important role in the Wilsonian
renormalization group equations.
We identify the poles of $D=2$ and $D=4$ as the quadratic and logarithmic divergences
after employing the dimensional regularization to take into account the quadratic divergence
following Refs \cite{Harada:2003jx}, denoted as:
\begin{align}
 A &\equiv \frac{\Lambda^2}{(4\pi)^2},\nonumber\\
 B &\equiv \frac{1}{(4\pi)^2}\ln\Lambda^2.
\end{align}
We evaluate the divergent part using $A$, $B$ as follows
\begin{align}
 A(m^2 )|_\text{div}  &= A - m^2 B, \nonumber\\ 
  B  (p,m_1 ^2 ,m_2 ^2 )|_\text{div} & =B,\nonumber\\
 B^\mu  (p,m_1 ^2 ,m_2 ^2 )|_\text{div} & = \frac{{p^\mu  }}
{2}B, \nonumber\\ 
 B^{\mu \nu } (p,m_1 ^2 ,m_2 ^2 )|_\text{div}  &=  - g^{\mu \nu } (2A - (m_1 ^2  + m_2 ^2 )B) - P^{\mu \nu }\frac{B}
{3}, \nonumber\\ 
 \tilde B^{\mu \nu } (p,m_1 ^2 ,m_2 ^2 )|_\text{div}  &= \frac{1}
{4}(B^{\mu\nu}+p^\mu p^\nu B)\nonumber\\
&=\frac{1}{4}g^{\mu \nu } \left( - 2A + (m_1 ^2  + m_2 ^2 )B\right) - P^{\mu \nu }\frac{B}
{12} + \frac{1}{4}p^\mu p^\nu B,
\label{divergentpart}
\end{align}
where $P^{\mu\nu}\equiv p^2g^{\mu\nu}-p^\mu p^\nu$.

We take the Landau gauge ($\alpha_V=\alpha_A=0$) in the following 
calculations of one-loop corrections.
\subsection{$\mathcal{A}-\mathcal{A}$}
\begin{figure}
\begin{align}
\delta^{ab}\Pi_{\mathcal{AA}}^{\mu\nu}&= \phs{$\pi$}+\phss{$\sigma$}{$\pi$}+\phss{$p$}{$\sigma$}+\nonumber\\
&+\phvs{$A_1$}{$\sigma$}+\phvs{$\rho$}{$p$}
+\phvs{$\rho$}{$\pi$} .\nonumber
\end{align}
\caption{Contributions to $\Pi_{\mathcal{AA}}^{\mu\nu}$ at one-loop. }
\label{fig:AA}
\end{figure}
The contributions from each diagram shown in Fig. \ref{fig:AA} are
\begin{align}
(1): \scalebox{1.0}{\phs{$\pi$}}~~~&=  - (1 + \beta \gamma ^2  - a)N_f \delta ^{ab} g^{\mu \nu } A(0), 
 \displaybreak[0]\\ 
(2):\scalebox{1.0}{\phvs{$\rho$}{$\pi$}} & = -g^2 F_\pi  ^2 (a - \beta \gamma ^2 )^2 N_f \delta ^{ab} g^{\mu \nu } B(p,0,M_\rho  ^2 ), \displaybreak[0]\\ 
(3):\scalebox{1.0}{\phvs{$\rho$}{$p$}}&  = -g^2 F_\pi  ^2 \beta \gamma ^2 N_f \delta ^{ab}g^{\mu \nu } B(p,0,M_\rho  ^2 ), \displaybreak[0]\\ 
 (4):\scalebox{1.0}{\phvs{$A_1$}{$\sigma$}} &= -\frac{{F_\pi  ^2 }}
{a}\beta ^2 \gamma ^2 g^2 N_f \delta ^{ab}g^{\mu \nu } B(p,0,M_{A_1 } ^2 ), \displaybreak[0]\\ 
(5):\scalebox{1.0}{\phss{$p$}{$\sigma$}}  &  = \frac{\beta }
{a}\gamma ^2 N_f \delta ^{ab} \tilde B^{\mu \nu } (p, 0, 0 ), \displaybreak[0]\\ 
(6):\scalebox{1.0}{\phss{$\sigma$}{$\pi$}} &  = \frac{1}
{a}N_f \delta ^{ab} \Big[(\beta \gamma ^2  - a)^2 \tilde B^{\mu \nu } (p, 0, 
 0) - \beta \gamma ^2 (\beta \gamma ^2  - a)(B^\mu  (p, 0, 0 )p^\nu   + 
 p^\mu  B^\nu  (p, 0, 0) )\nonumber\\ 
 &~~~~~~~~~~~~~~~  + \beta ^2 \gamma ^4 p^\mu  p^\nu  B(p,0,0 )\Big].
\end{align} 
With Eq.(\ref{divergentpart}) the divergent parts are given by
\begin{align}
 & (1): - N_f \delta ^{ab} g^{\mu\nu}(1 + \beta \gamma ^2  - a)A ,\nonumber\\ 
 & (2):- N_f \delta ^{ab} g^{\mu\nu}g^2 F_\pi  ^2 (a - \beta \gamma ^2 )^2 \frac{3}{4}B,\nonumber\\ 
 & (3):-N_f \delta ^{ab} g^{\mu\nu}g^2 F_\pi  ^2 \beta \gamma ^2 \frac{{3}}{4}B,\nonumber\\ 
 & (4):-N_f \delta ^{ab} g^{\mu\nu}g^2F_\pi  ^2\frac{\beta ^2\gamma ^2}{a}  \frac{3}{4}B,\nonumber\\ 
 & (5): \frac{\beta}{a}\gamma ^2 N_f \delta ^{ab}\frac{1}
{4}\left( - 2g^{\mu \nu } A - P^{\mu \nu } \frac{B}
{3} + p^\mu  p^\nu  B\right) ,\nonumber\\ 
 & (6): \frac{1}{a}N_f \delta ^{ab} \left((\beta \gamma ^2  - a)^2 \frac{1}
{4}\left( - 2g^{\mu \nu } A - P^{\mu \nu } \frac{B}
{3}\right) + \frac{1}{4}(\beta \gamma ^2  + a)^2 p^\mu  p^\nu  B\right).
\end{align}
The sum of divergent parts is given by
\begin{align}
\left.\Pi_{\mathcal{AA}}^{\mu\nu}\right|_\text{div}
&=N_f\left[ g^{\mu\nu}\left(A\left(  -(1 + \beta \gamma ^2  - a)
-\frac{\beta }{2a}\gamma ^2  - \frac{1}{{2a}}(\beta \gamma ^2  - a)^2 \right)\right.
- \frac{3}{4}g^2 F_\pi  ^2B\left( (a - \beta \gamma ^2 )^2 +\beta \gamma ^2
   +\frac{1}{a}\beta ^2 \gamma ^2 \right)\right)\nonumber\\ 
&~~~~~~~~~~ -\frac{{P^{\mu \nu } }}
{{12}}B\left(\frac{\beta}{a} \gamma ^2  + \frac{1}{a}(a - \beta \gamma ^2 )^2 \right)
+  \left.\frac{1}{4}p^\mu  p^\nu  B\left(\frac{\beta }{a}\gamma ^2  + \frac{1}{a}(\beta \gamma ^2  + a)^2 \right)\right].
\end{align}
The divergences can be canceled by the following counterterms:
\begin{align}
\left.\Pi_{\mathcal{AA}}^{\mu\nu}\right|_\text{div}
&=-(\delta_{F_\pi^2}+\delta_{F_{A_1}^2})g^{\mu\nu}-2\delta_{Z_2}P^{\mu\nu}
+\delta_{\cal{A}\cal{A}}p^\mu p^\nu,
\label{eq:cntrAA}
\end{align}
where we introduce the counterterm $\delta_{\cal{A}\cal{A}}$ to
cancel the divergent part proportional to $p^\mu p^\nu$.
Such counterterm  does not exist in the 
Lagrangian of any order. This divergence looks accidental. 
However this divergence is automatically canceled when one calculates an observable such as $\langle T{\cal A}_\mu(x){\cal A}_\nu(0)\rangle$.
Therefore one may simply drop this divergence.
\subsection{$\mathcal{V}-\mathcal{V}$}
\begin{figure}
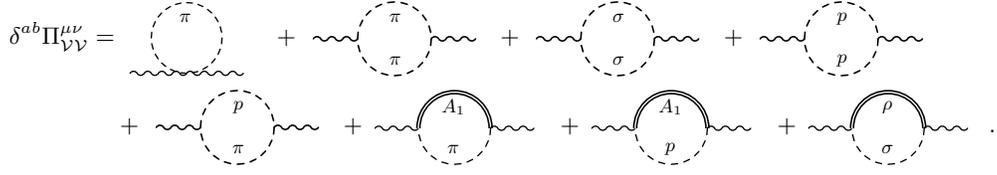

 \begin{align}
\delta^{ab}\Pi_{\mathcal{VV}}^{\mu\nu}=& \phs{$\pi$}+\phss{$\pi$}{$\pi$}+\phss{$\sigma$}{$\sigma$}+\phss{$p$}{$p$}\nonumber\\
&+\phss{$p$}{$\pi$}+\phvs{$A_1$}{$\pi$}+\phvs{$A_1$}{$p$}
+\phvs{$\rho$}{$\sigma$} .\nonumber
\end{align}
\caption{Contributions to $\Pi_{\mathcal{VV}}^{\mu\nu}$ at one-loop. }
\label{fig:VV}
\end{figure}

The contributions from each diagram shown in Fig. \ref{fig:VV} are
 \begin{align}
(1):\scalebox{1.0}{ ~~\phs{$\pi$}} ~~
&  = (1 + \beta \gamma ^2  - a)N_f \delta ^{ab} 
  {{g}}^{\mu \nu } A(0) ,
 \displaybreak[0]\\ 
(2):\scalebox{1.0}{\phss{$\pi$}{$\pi$}}
 &  = \frac{1}
{8}(a(1 - \gamma ^2 ) - 2)^2 N_f \delta ^{ab} B^{\mu \nu } (p,0,0) ,
\displaybreak[0]\\ 
(3):\scalebox{1.0}{\phss{$\sigma$}{$\sigma$}}
 &  = \frac{1}
{8}N_f \delta ^{ab} B^{\mu \nu } (p, 0, 0),
\displaybreak[0]\\ 
(4):\scalebox{1.0}{\phss{$p$}{$p$}}
 &  = \frac{{a^2 }}
{{8\beta ^2 }}N_f \delta ^{ab} B^{\mu \nu } (p, 0, 0) ,
\displaybreak[0]\\ 
(5):\scalebox{1.0}{\phss{$p$}{$\pi$}}
 &  = \frac{{\gamma ^2 }}
{{4\beta }}N_f \delta ^{ab} ((a - \beta )^2 B^{\mu \nu } (p, 0, 0) + 
  p^\mu  p^\nu \beta^2 B(p, 0, 0)\nonumber\\
 &~~~+ 2(a - \beta )\beta (2p^\mu  B^\nu  (p, 0, 0 ) -   
  p^\mu  p^\nu  B(p, 0, 0 ))) ,
 \displaybreak[0]\\ 
(6):\scalebox{1.0}{\phvs{$A_1$}{$p$}}
 &  = -\frac{{F_\pi  ^2 a^2 {{g}}^{{2}} }}
{\beta }N_f \delta ^{ab} g^{\mu \nu } B(p, 0,M_{A_1 } ^2 ),
\displaybreak[0]\\ 
(7):\scalebox{1.0}{\phvs{$A_1$}{$\pi$}}
 &  = -F_\pi  ^2 g^2 \gamma ^2 (\beta  - a)^2 N_f \delta ^{ab}g^{\mu \nu } B(p,0,M_{A_1 } ^2 ), \displaybreak[0]\\
(8):\scalebox{1.0}{\phvs{$\rho$}{$\sigma$}}
 &  = -aF_\pi  ^{{2}} {{g}}^{{2}} N_f \delta ^{ab}g^{\mu \nu } B(p,0 ,M_\rho  ^2 ).\displaybreak[0]
\end{align}
The divergent parts are
\begin{align}
(1):&(1 + \beta \gamma ^2  - a)N_f \delta ^{ab} g^{\mu \nu } A, \nonumber\displaybreak[0]\\ 
(2):&\frac{1}{8}(a(1 - \gamma ^2 ) - 2)^2 N_f \delta ^{ab} 
\left( - 2g^{\mu \nu } A - P^{\mu \nu } \frac{B}{3}\right),\nonumber\displaybreak[0]\\ 
(3):&\frac{1}{8}N_f \delta ^{ab} 
\left( - 2g^{\mu \nu } A - P^{\mu \nu } \frac{B}{3}\right),\nonumber\displaybreak[0]\\
(4):& \frac{{a^2 }}
{{8\beta ^2 }}N_f \delta ^{ab} \left(-2g^{\mu \nu } A - P^{\mu \nu } \frac{B}{3}\right),\nonumber\displaybreak[0]\\ 
(5):&\frac{{\gamma ^2 }}{{4\beta }}N_f \delta ^{ab} \left((a - \beta )^2 \left(  - 2Ag^{\mu \nu }  - P^{\mu \nu } \frac{B}{3}\right)  + \beta ^2 p^\mu  p^\nu  B\right),\nonumber\displaybreak[0]\\ 
(6):& - \frac{{F_\pi  ^2 a^2 g^2 }}
{\beta }N_f \delta ^{ab} g^{\mu \nu } \frac{3}
{4}B,\nonumber\displaybreak[0]\\ 
(7):&- F_\pi  ^2 g^2 \gamma ^2 (\beta  - a)^2 N_f \delta ^{ab} g^{\mu \nu } \frac{3}
{4}B ,\nonumber\displaybreak[0]\\ 
(8):&- aF_\pi  ^2 g^2 N_f \delta ^{ab} g^{\mu \nu } \frac{3}{4}B.
\end{align}
The sum of divergent parts is given by
\begin{align}
\left.\Pi_{\mathcal{VV}}^{\mu\nu}\right|_\text{div}
&= N_f\left[g^{\mu \nu }\left(A\left( (1 + \beta \gamma ^2  - a) + \frac{{ - 1}}
{4}(a(1 - \gamma ^2 ) - 2)^2  - \frac{1}{4} - \frac{{a^2 }}{{4\beta ^2 }} - \frac{{\gamma ^2 }}
{{2\beta }}(a - \beta )^2 )\right) \right.\right.\nonumber\\
 &~~~~~~~~~ - \left.\frac{3}{4}F_\pi  ^2 B\left( \frac{{a^2 g^2 }}{\beta } + g^2 \gamma ^2 (\beta  - a)^2  + ag^2 \right)\right) \nonumber\\
&~~~~~~~~~  - \frac{{P^{\mu \nu } }}{{24}}B((a(1 - \gamma ^2 ) - 2)^2  + 1 + \frac{{a^2 }}
{{\beta ^2 }} + \frac{{2\gamma ^2 }}{\beta }(a - \beta )^2 )
+ p^\mu  p^\nu  B\left.\frac{\beta\gamma ^2 }{4}\right].
\end{align}
The divergent part is canceled by following counter terms:
\begin{align}
\left.\Pi_{\mathcal{VV}}^{\mu\nu}\right|_\text{div} &=-\delta_{F_\sigma^2}g^{\mu\nu}
-2\delta_{z1}P^{\mu\nu},
\label{eq:cntrVV}
\end{align}
where we drop the divergent term proportional to $p^\mu p^\nu$.
\subsection{$\mathcal{A}-A_1$}
\begin{figure}
 \begin{align}
\delta^{ab}\Pi_{\mathcal{A}A_1}^{\mu\nu}&= \pvs{$\pi$}+\pvs{$p$}+\pvs{$\sigma$}+
\pvss{$\sigma$}{$\pi$}\nonumber\\
&+\pvss{$\sigma$}{$p$}+\pvvs{$\rho$}{$p$}
+\pvvs{$\rho$}{$\pi$} .\nonumber
\end{align}
\caption{Contributions to $\Pi_{\mathcal{A}A_1}^{\mu\nu}$ at one-loop. }
\label{fig:AA1}
\end{figure}
The contributions from each diagram shown in Fig. \ref{fig:AA1} are

 \begin{align}
(1):~~~\scalebox{1.0}{\pvs{$\pi$}}
 &  = \frac{{ - g\gamma }}
{2}g^{\mu \nu } (2a - \beta  - \beta \gamma ^2 )N_f \delta ^{ab} A(0) ,\displaybreak[0]\\ 
(2):~~~\scalebox{1.0}{\pvs{$p$}}
 &  = \frac{{g\gamma }}
{2}g^{\mu \nu } N_f \delta ^{ab} A(0) ,\displaybreak[0]\\ 
(3):~~~\scalebox{1.0}{\pvs{$\sigma$}}
 &  = \frac{{g\gamma }}
{2}\frac{\beta }
{a}g^{\mu \nu } N_f \delta ^{ab} A(0) ,\displaybreak[0]\\ 
(4):\scalebox{1.0}{\pvss{$\sigma$}{$\pi$}}
 &  = \frac{{g\gamma }}{{2a}}\frac{1}
{4}N_f \delta ^{ab} \left(2(a - \beta )(\beta \gamma ^2  - a)B^{\mu \nu } (p,0,0) - 2(a + \beta \gamma ^2 )(a - \beta \gamma )p^\mu  p^\nu  B(p,0,0)\right. \nonumber\\ 
 &~~~  + 4(\beta \gamma ^2  - a)(a - \beta \gamma )p^\nu  B^\mu  (p,0,0) - 4(a - \beta )(a + \beta \gamma ^2 )p^\mu  B^\nu  (p,0,0) \nonumber\\ 
 &~~~  \left.- 2\left((\beta \gamma ^2  - a)(a - \beta \gamma ) - (a - \beta )(a + \beta \gamma ^2 )\right)p^\mu  p^\nu  B(p,0,0)\right) ,\displaybreak[0]\\ 
(5):\scalebox{1.0}{\pvss{$\sigma$}{$p$}}
 &  = \frac{{g\gamma }}
{{2a}}N_f \delta ^{ab} \frac{1}{4}\left(2a B^{\mu \nu } (p,0,0) - 2(a - \beta \gamma )p^\mu  p^\nu  B(p,0,0) \right.\nonumber\\ 
 &~~~  + \left.4(a - \beta \gamma )p^\nu  B^\mu  (p,0,0) - 4a p^\mu  B^\nu  (p,0,0) + 2\beta \gamma p^\mu  p^\nu  B(p,0,0)\right) \displaybreak[0],\\ 
(6):\scalebox{1.0}{\pvvs{$\rho$}{$p$}}
 &  = a F_\pi  ^2 \gamma (\beta  - a)g^2 N_f \delta ^{ab} \left(g^{\mu \nu } B(p,0,M_\rho  ^2 )
 - \frac{1}
{M_\rho^2 }\left(\widetilde B^{\mu \nu } (p,0,M_\rho  ^2 ) - \widetilde B^{\mu \nu } (p,0,0)\right)\right) ,\displaybreak[0]\\ 
(7):\scalebox{1.0}{\pvvs{$\rho$}{$\pi$}}
 &  =  - g\gamma F_\pi  ^2 (a - \beta \gamma ^2 )(\beta  - a)g^2 N_f \delta ^{ab} 
\left(g^{\mu \nu } B(p,0,M_\rho  ^2 ) - \frac{1}
{{M_\rho  ^2 }}\left(\widetilde B^{\mu \nu } (p,0,M_\rho  ^2 ) - \widetilde B^{\mu \nu } (p,0,0)\right)\right).
\end{align}
The divergent parts are given by
\begin{align}
(1): &\frac{{ - g\gamma }}
{2}g^{\mu \nu } (2a - \beta  - \beta \gamma ^2 )N_f \delta ^{ab} A ,\nonumber\\ 
(2):&\frac{{g\gamma }}
{2}g^{\mu \nu } N_f \delta ^{ab} A ,\nonumber\\ 
(3):&\frac{{g\gamma }}
{2}\frac{\beta }
{a}g^{\mu \nu } N_f \delta ^{ab} A ,\nonumber\\ 
(4):&\frac{{g\gamma }}{{4a}}N_f \delta ^{ab} \left((a - \beta )(\beta \gamma ^2  - a)
\left( - 2g^{\mu \nu } A - P^{\mu \nu } \frac{B}{3}\right) - (a + \beta \gamma ^2 )(a - \beta \gamma )p^\mu  p^\nu  B\right),\nonumber\\ 
(5):&\frac{{g\gamma }}{{2a}}N_f \delta ^{ab} \frac{1}
{4}\left(2a\left( - 2g^{\mu \nu } A - P^{\mu \nu } \frac{B}
{3}\right) - 2(a - \beta \gamma )p^\mu  p^\nu  B\right),\nonumber\\ 
(6):&g F_\pi  ^2 \gamma (\beta  - a)g^2 g^{\mu \nu } N_f \delta ^{ab} B\frac{3}{4},\nonumber\\
(7):& - g\gamma F_\pi  ^2 (a - \beta \gamma ^2 )(\beta  - a)g^2 g^{\mu \nu } N_f \delta ^{ab} B\frac{3}{4} .
\end{align}
The sum of divergent parts is
\begin{align}
\left.\Pi_{\mathcal{A}A_1}^{\mu\nu}\right|_\text{div}&= N_f g\gamma \left[g_{\mu\nu}\left(\frac{A}{2}\left(  - (2a - \beta  - \beta \gamma ^2 ) + 1 + \frac{\beta }
{a} - \frac{1}
{a}(a - \beta )(\beta \gamma ^2  - a) - 1\right)\right.\right.  \nonumber\\ 
 &~~~~~~~~~~~~~  + \left.B\frac{3}
{4}F_\pi  ^2 g^2 \left( (\beta  - a) - (a - \beta \gamma ^2 )(\beta  - a)\right)\right) \nonumber\\ 
&~~~~~~~~~~~~~\left.- P^{\mu \nu } \frac{B}{{12}}\left(\frac{1}{a}(a - \beta )(\beta \gamma ^2  - a) + 1\right)
- p^\mu  p^\nu  B\frac{1}{4a}(a - \beta \gamma )\left(a + \beta \gamma ^2  + 1\right)\right].
\end{align}  
The divergences are canceled out by the following counterterms:
\begin{align}
\left.\Pi_{\mathcal{A}A_1}^{\mu\nu}\right|_\text{div}
&=g^{\mu\nu}\frac{g F_p F_{A_1}}{2}\left( \frac{\delta_{F_{A_1}^2}}{F_{A_1}^2}
+\delta_{Z_{A1}}\frac{\delta_{F_p^2}}{F_p^2}+\delta_{Z_g}\right)-2g\delta_{z_4}P^{\mu\nu}.
\label{eq:cntrAA_1}
\end{align}
where we drop the divergent term proportional to $p^\mu p^\nu$.
\subsection{$\mathcal{V}-\rho$}
\begin{figure}
 \begin{align}
\delta^{ab}\Pi_{\mathcal{V}\rho}^{\mu\nu}&= \pvs{$\pi$}+\pvs{$p$}+\pvs{$\sigma$}+
\pvss{$\pi$}{$\pi$}+\pvss{$\sigma$}{$\sigma$}\nonumber\\
&~~+\pvss{$p$}{$p$}+\pvss{$p$}{$\pi$}+\pvvs{$A_1$}{$\pi$}
+\pvvs{$A_1$}{$p$}+\pvvs{$\rho$}{$\sigma$}.\nonumber
\end{align}
\caption{Contributions to $\Pi_{\mathcal{V}\rho}^{\mu\nu}$ at one-loop. }
\label{fig:Vrho}
\end{figure}

The contributions from each diagram shown in Fig. \ref{fig:Vrho} are
 \begin{align}
(1):~~~\scalebox{1.0}{\pvs{$\pi$}}
 &  =  - \frac{1}{2}g(2\beta \gamma ^2  - a\gamma ^2  - a)N_f \delta ^{ab} g^{\mu \nu } 
  A(0), \displaybreak[0]\\ 
(2):~~~\scalebox{1.0}{\pvs{$p$}}
 &  = \frac{1}
{2}g\frac{a}{\beta }N_f \delta ^{ab} g^{\mu \nu } A(0), \displaybreak[0]\\ 
(3):~~~\scalebox{1.0}{\pvs{$\sigma$}}
 &  = \frac{1}{2}gN_f \delta ^{ab} g^{\mu \nu } A(0) ,\displaybreak[0]\\ 
(4):\scalebox{1.0}{\pvss{$\pi$}{$\pi$}}
 &  = \frac{1}
{8}ag(1 - \gamma ^2 )(2 - a(1 - \gamma ^2 ))N_f \delta ^{ab} B^{\mu \nu } (p,0,0)
\displaybreak[0],\\ 
(5):\scalebox{1.0}{\pvss{$\sigma$}{$\sigma$}}
 &  = \frac{1}
{8}gN_f \delta ^{ab} B^{\mu \nu } (p,0,0) ,\displaybreak[0]\\ 
(6):\scalebox{1.0}{\pvss{$p$}{$p$}}
 &  = \frac{{(2\beta  - a)}}
{{8\beta ^2 }}agN_f \delta ^{ab} B^{\mu \nu } (p,0,0) ,\displaybreak[0]\\ 
(7):\scalebox{1.0}{\pvss{$p$}{$\pi$} }
 &  = \frac{{ - \gamma g}}
{{4\beta }}N_f \delta ^{ab} (\gamma (a - \beta )^2 B^{\mu \nu } (p,0,0) - \beta (a - \beta \gamma )p^\mu  p^\nu  B(p,0,0) \nonumber\\ 
 &~~~  - 2\beta \gamma (a - \beta )p^\mu  B^\nu   + 2(a - \beta )(a - \beta \gamma )p^\nu  B^\mu   - (a - \beta )(a - 2\beta \gamma )p^\mu  p^\nu  B) ,\displaybreak[0]\\ 
(8):\scalebox{1.0}{\pvvs{$A_1$}{$\pi$}}
 &   = F_\pi  ^2 g^3 \gamma ^2 (\beta  - a)^2 N_f \delta ^{ab} \left( {g^{\mu \nu } B(p,0,M_{A_1 }^2 ) - \frac{1}
{{M_{A_1 } ^2 }}\left( {\tilde B^{\mu \nu } (p,0,M_{A_1 } ^2 ) - \tilde B^{\mu \nu } (p,0,0)} \right)} \right) ,\displaybreak[0]\\ 
(9):\scalebox{1.0}{\pvvs{$A_1$}{$p$}}
 &  =  - F_\pi  ^2g^3 \frac{a}
{\beta }(\beta  - a) N_f \delta ^{ab} \left( {g^{\mu \nu } B(p,0,M_{A_1 }^2 ) - \frac{1}
{{M_{A_1 } ^2 }}\left( {\tilde B^{\mu \nu } (p,0,M_{A_1 } ^2 ) - \tilde B^{\mu \nu } (p,0,0)} \right)} \right),\displaybreak[0]\\ 
(10):\scalebox{1.0} {\pvvs{$\rho$}{$\sigma$}}
&=0.
\end{align}
The divergent parts are given by
\begin{align}
(1):& - \frac{1}{2}g(2\beta \gamma ^2  - a\gamma ^2  - a)N_f \delta ^{ab} g^{\mu \nu } A,\nonumber\displaybreak[0]\\
(2):&\frac{1}{2}g\frac{a}{\beta }N_f \delta ^{ab} g^{\mu \nu } A,\nonumber\displaybreak[0]\\
(3):&\frac{1}{2}gN_f \delta ^{ab} g^{\mu \nu } A,\nonumber\displaybreak[0]\\
(4):&\frac{1}
{8}ag(1 - \gamma ^2 )(2 - a(1 - \gamma ^2 ))N_f \delta ^{ab} \left( { - 2g^{\mu \nu } A - P^{\mu \nu } \frac{B}{3}} \right),\nonumber\displaybreak[0]\\
(5):&\frac{1}
{8}gN_f \delta ^{ab} \left( { - 2g^{\mu \nu } A - P^{\mu \nu } \frac{B}
{3}} \right),\nonumber\displaybreak[0]\\
(6):&\frac{{(2\beta  - a)}}
{{8\beta ^2 }}agN_f \delta ^{ab} \left( { - 2g^{\mu \nu } A - P^{\mu \nu } \frac{B}
{3}} \right),\nonumber\displaybreak[0]\\
(7):&\frac{{ - \gamma g}}
{{4\beta }}N_f \delta ^{ab} \left( {\gamma (a - \beta )^2 \left( { - 2g^{\mu \nu } A - P^{\mu \nu } \frac{B}
{3}} \right) + \beta (\beta \gamma  - a)p^\mu  p^\nu  B} \right),\nonumber\displaybreak[0]\\
(8):&gN_f \delta ^{ab} g^{\mu \nu } \gamma ^2 (\beta  - a)^2 \frac{3}{4}F_\pi  ^2 g^2 B,\nonumber\displaybreak[0]\\
(9):& - g \frac{a}
{\beta }(\beta  - a) N_f \delta ^{ab} F_\pi  ^2g^2\frac{3}{4}B,\nonumber\displaybreak[0]\\
(10):&0.
\end{align}
The sum of divergent parts is given by
\begin{align}
\Pi_{\mathcal{V}\rho}^{\mu\nu}&=
 gN_f \left[g^{\mu \nu }\left(A\left(\frac{1}{2} - \beta \gamma ^2  + \frac{1}
{2}a\gamma ^2  + \frac{1}
{2}a + \frac{a}
{{2\beta }} + \frac{{ - 1}}
{4}a(1 - \gamma ^2 )(2 - a(1 - \gamma ^2 ))\right.\right.\right.\nonumber\\
  &~~~~~~~~~~~-\left.\frac{1}{4} - \frac{{(2\beta  - a)}}
{{4\beta ^2 }}a + \frac{{\gamma ^2 }}
{{2\beta }}(a - \beta )^2 \right)+ \left.\frac{3}{4}F_\pi  ^2 g^2 B\left(\gamma ^2 (\beta  - a)^2  - \frac{a}
{\beta }(\beta  - a)\right)\right)\nonumber\\
&~~~~~~~~~~~ -P^{\mu\nu}\frac{B}{24}\left(a(1 - \gamma ^2 )(2 - a(1 - \gamma ^2 )) + 1 + \frac{{(2\beta  - a)}}
{{\beta ^2 }}a + \frac{{ - 2\gamma ^2 }}
{\beta }(a - \beta )^2 \right)
\left.- p^\mu  p^\nu  B\frac{{\gamma}}
{4}(\beta \gamma  - a)\right].
\end{align}  
The divergence is canceled  by the following counterterms:
\begin{align}
\left. {\Pi _{\mathcal{V}\rho }^{\mu \nu } } \right|_{{\text{div}}}  = \frac{{F_\sigma  ^2 g}}
{2}\left( {2\frac{{\delta _{F_\sigma  ^2 } }}
{{F_\sigma  ^2 }} + \delta _{Z_g }  + \delta _{Z_\rho  } } \right)g^{\mu \nu }  - 2g\delta _{Z_4 } P^{\mu \nu }.
\label{eq:cntrVrho}
\end{align}
where we drop the divergent term proportional to $p^\mu p^\nu$.
\subsection{$\rho-\rho$}
\begin{figure}
 \begin{align}
\delta^{ab}\Pi_{\rho\rho}^{\mu\nu}&=\vvs{$\pi$}+\vvs{$p$}+\vvs{$\sigma$}+\vvv{$\rho$}\nonumber\\
&+\vvv{$A_1$}+\vvss{$\pi$}{$\pi$}+\vvss{$p$}{$p$}+\vvss{$\sigma$}{$\sigma$}\nonumber\\
&+\vvss{$p$}{$\pi$}+\vvvs{$A_1$}{$\pi$}+\vvvs{$\rho$}{$\sigma$}+\vvvs{$A_1$}{$p$}\nonumber\\
&+\vvvv{$\rho$}{$\rho$}+\vvvv{$A_1$}{$A_1$}
+\vvgg{$C_V$}{$C_V$}+\vvgg{$C_A$}{$C_A$} .\nonumber
\end{align}
\caption{Contributions to $\Pi_{\rho\rho}^{\mu\nu}$ at one-loop. }
\label{fig:rhorho}
\end{figure}

The contributions from each diagram shown in Fig. \ref{fig:rhorho} are
 \begin{align}
(1):\scalebox{1.0}{\vvs{$\pi$}}
 &  =  - g^2 (a - \beta )\gamma ^2 N_f \delta ^{ab} g^{\mu \nu } A(0) ,\\ 
(2):\scalebox{1.0}{\vvs{$p$}}
 &  = \frac{{ - g^2 }}
{\beta }(a - \beta )N_f \delta ^{ab} g^{\mu \nu } A(0) ,\displaybreak[0]\\ 
(3):\scalebox{1.0}{\vvs{$\sigma$}}
&=0,\displaybreak[0]\\
(4):\scalebox{1.0}{\vvv{$\rho$}}
 &  = g^2 N_f \delta ^{ab} g^{\mu \nu } \frac{{(D - 1)^2 }}
{D}A(M_\rho  ^2 ) ,\displaybreak[0]\\
(5):\scalebox{1.0}{\vvv{$A_1$}}
 &  = g^2 N_f \delta ^{ab} g^{\mu \nu } \frac{{(D - 1)^2 }}
{D}A(M_{A_1 } ^2 ) \displaybreak[0],\\ 
(6):\scalebox{1.0}{\vvss{$\pi$}{$\pi$}}
 &  = \frac{1}
{8}a^2 g^2 (1 - \gamma ^2 )^2N_f \delta ^{ab} B^{\mu \nu } (p,0,0) ,\displaybreak[0]\\ 
(7):\scalebox{1.0}{\vvss{$p$}{$p$}}
 &  = \frac{1}
{{8\beta ^2 }}(2\beta  - a)^2 g^2 N_f \delta ^{ab} B^{\mu \nu } (p,0,0) ,\displaybreak[0]\\ 
(8):\scalebox{1.0}{\vvss{$\sigma$}{$\sigma$}}
 &  = \frac{1}
{8}g^2 B^{\mu \nu } (p,0,0) ,\displaybreak[0]\\ 
(9):\scalebox{1.0}{\vvss{$p$}{$\pi$}}
 &  = \frac{1}
{{4\beta }}g^2 N_f \delta ^{ab} (\gamma ^2 (a - \beta )^2 B^{\mu \nu } (p,0,0) + (\beta \gamma  - a)^2 p^\mu  p^\nu  B(p,0,0) \displaybreak[0]\nonumber\\ 
 &~~~  + 2\gamma (\beta  - a)(\beta \gamma  - a)( p^\mu  B^\nu  (p,0,0) + p^\nu  B^\mu  (p,0,0) - p^\mu  p^\nu  B(p,0,0)) ) ,\displaybreak[0]\\ 
(10):\scalebox{1.0}{\vvvs{$A_1$}{$\pi$}}
 &=  - F_\pi  ^2 (\beta  - a)^2 \gamma ^2 g^4N_f \delta ^{ab} \left( {g^{\mu \nu } B(p,0,M_{A_1 }^2 ) - \frac{1}
{{M_{A_1 } ^2 }}\left( {\tilde B^{\mu \nu } (p,0,M_{A_1 } ^2 ) - \tilde B^{\mu \nu } (p,0,0)} \right)} \right),\\
(11):\scalebox{1.0}{\vvvs{$\rho$}{$\sigma$}}
&=0,\\ \displaybreak[0]\nonumber\\
(12):\scalebox{1.0}{\vvvs{$A_1$}{$p$}}
& =  - \frac{{F_\pi  ^2 }}
{\beta }(\beta  - a)^2 g^4 N_f \delta ^{ab} \left( {g^{\mu \nu } B(p,0,M_{A_1 }^2 ) - \frac{1}
{{M_{A_1 } ^2 }}\left( {\tilde B^{\mu \nu } (p,0,M_{A_1 } ^2 ) - \tilde B^{\mu \nu } (p,0,0)} \right)} \right)\displaybreak[0],\\
(13):\scalebox{1.0}{\vvvv{$\rho$}{$\rho$}}
 &  = \frac{1}
{2}g^2 N_f \delta ^{ab} {\int\frac{\textrm{d}^4k}{i(2\pi)^4}}\frac{1}
{{((k - p)^2  - M_\rho  ^2 )(k^2  - M_\rho  ^2 )}} \nonumber\\ 
 &~~~  \times \{  - 2p^\rho  g^{\mu \sigma }  + 2p^\sigma  g^{\mu \rho }  + (p - 2k)^\mu  g^{\rho \sigma } \} \{  - 2p^{\rho '} g^{\nu \sigma '}  + 2p^{\sigma '} g^{\nu \rho '}  + (p - 2k)^\nu  g^{\rho '\sigma '} \}  \nonumber\\ 
 &~~~  \times (g_{\rho \rho '}  - \frac{{k_\rho  k_{\rho '} }}
{{k^2 }})(g_{\sigma \sigma '}  - \frac{{(k - p)_\sigma  (k - p)_{\sigma '} }}
{{(k - p)^2 }}),\displaybreak[0]\\
(14):\scalebox{1.0}{\vvvv{$A_1$}{$A_1$}}
 &  = \frac{1}
{2}g^2 N_f \delta ^{ab} {\int\frac{\textrm{d}^4k}{i(2\pi)^4}}\frac{1}
{{((k - p)^2  - M_{A_1 } ^2 )(k^2  - M_{A_1 } ^2 )}} \nonumber\\ 
 &~~~  \times \{  - 2p^\rho  g^{\mu \sigma }  + 2p^\sigma  g^{\mu \rho }  + (p - 2k)^\mu  g^{\rho \sigma } \} \{  - 2p^{\rho '} g^{\nu \sigma '}  + 2p^{\sigma '} g^{\nu \rho '}  + (p - 2k)^\nu  g^{\rho '\sigma '} \}  \nonumber\\ 
 &~~~  \times (g_{\rho \rho '}  - \frac{{k_\rho  k_{\rho '} }}
{{k^2 }})(g_{\sigma \sigma '}  - \frac{{(k - p)_\sigma  (k - p)_{\sigma '} }}
{{(k - p)^2 }}),\\
(15):\scalebox{1.0}{\vvgg{$C_V$}{$C_V$}}
 &  = \frac{{ - g^2 }}
{4}N_f \delta ^{ab} (B^{\mu \nu } (p,0,0) - p^\mu  p^\nu  B(p,0,0)) ,\\ 
(16):\scalebox{1.0}{\vvgg{$C_A$}{$C_A$}}
 &  = \frac{{ - g^2 }}
{4}N_f \delta ^{ab} (B^{\mu \nu } (p,0,0) - p^\mu  p^\nu  B(p,0,0)),
\end{align}
where $D$ is the space-time dimension. 
To keep the gauge invariance we take $D\to4$ after the sum of diagram.

The divergent parts are given by
\begin{align}
 & (1): - g^2 (a - \beta )\gamma ^2 N_f \delta ^{ab} g^{\mu \nu } A \nonumber,\displaybreak[0]\\ 
 & (2):\frac{{ - g^2 }}
{\beta }(a - \beta )N_f \delta ^{ab} g^{\mu \nu } A \nonumber,\displaybreak[0]\\ 
 & (3):0 \nonumber,\displaybreak[0]\\ 
 & (4):g^2 N_f \delta ^{ab} g^{\mu \nu } \frac{{(D - 1)^2 }}
{D}A(M_\rho^2)   \nonumber,\displaybreak[0]\\ 
 & (5):g^2 N_f \delta ^{ab} g^{\mu \nu } \frac{{(D - 1)^2 }}
{D}A({M_{A_1}^2})  \nonumber,\displaybreak[0]\\ 
 & (6):\frac{1}{8}a^2 g^2 (1 - \gamma ^2 )^2 N_f \delta ^{ab} \left( { - 2g^{\mu \nu } A - P^{\mu \nu } \frac{B}{3}} \right) \nonumber,\displaybreak[0]\\ 
 & (7):\frac{1}
{{8\beta ^2 }}(2\beta  - a)^2 g^2 N_f \delta ^{ab} \left( { - 2g^{\mu \nu } A - P^{\mu \nu } \frac{B}
{3}} \right) \nonumber,\displaybreak[0]\\ 
 & (8):\frac{1}
{8}g^2 N_f \delta ^{ab} \left( { - 2g^{\mu \nu } A - P^{\mu \nu } \frac{B}
{3}} \right) \nonumber,\displaybreak[0]\\ 
 & (9):\frac{1}
{{4\beta }}g^2 N_f \delta ^{ab} \left( {g^{\mu \nu } ( - 2\gamma ^2 (a - \beta )^2 A + (\beta \gamma  - a)^2 p^2 B) - P^{\mu \nu } \left( {\frac{{\gamma ^2 (a - \beta )^2 }}
{3} + (\beta \gamma  - a)^2 } \right)B} \right)\nonumber,\displaybreak[0]\\
 & (10): - F_\pi  ^2 (\beta  - a)^2 \gamma ^2 g^2 g^2 N_f \delta ^{ab} g^{\mu \nu } \frac{3}
{4}B \nonumber,\displaybreak[0]\\ 
 & (11):0 \nonumber,\displaybreak[0]\\ 
 & (12): - \frac{{F_\pi  ^2 }}
{\beta }(\beta  - a)^2 g^2 g^2 N_f \delta ^{ab} g^{\mu \nu } \frac{3}
{4}B \nonumber,\displaybreak[0]\\ 
 & (13):\frac{1}
{2}g^2 N_f \delta ^{ab} \left( {P^{\mu \nu } \frac{{14}}
{3}B - 2(D - 1)g^{\mu \nu } A(M_\rho^2)   - \frac{{p^2 g^{\mu \nu } }}
{2}B} \right) \nonumber,\displaybreak[0]\\ 
 & (14):\frac{1}
{2}g^2 N_f \delta ^{ab} \left( {P^{\mu \nu } \frac{{14}}
{3}B - 2(D - 1)g^{\mu \nu } A_{M_{A_1}^2}  - \frac{{p^2 g^{\mu \nu } }}
{2}B} \right)\nonumber,\displaybreak[0]\\ 
 & (15): - \frac{{g^2 }}
{4}N_f \delta ^{ab} \left( { - g^{\mu \nu } (2A + p^2 B) + P^{\mu \nu } \frac{{2B}}
{3}} \right) \nonumber,\displaybreak[0]\\ 
 & (16): - \frac{{g^2 }}
{4}N_f \delta ^{ab} \left( { - g^{\mu \nu } (2A + p^2 B) + P^{\mu \nu } \frac{{2B}}
{3}} \right).
\end{align}

The divergent part including the internal line of $\rho$ is given by
\begin{align}
(4) + (13) + (15)
 &= g^2 N_f \delta ^{ab} \left[ g^{\mu \nu } \left( {\frac{{(D - 1)^2 }}
{D}A(M_\rho^2)   - (D - 1)A(M_\rho^2)   - \frac{{p^2 }}
{4}B + \frac{1}
{4}(2A + p^2 B)} \right)\right.\nonumber\\
&~~~~~~~~~~~~~~~~+\left.P^{\mu \nu } \left( {\frac{7}{3}B - \frac{1}
{4}\frac{{2B}}
{3}} \right) \right] \nonumber\\ 
 &  = g^2 N_f \delta ^{ab} \left[ {g^{\mu \nu } \left( {\frac{1}{D}(1 - \frac{1}
{2}D)A(M_\rho^2)   - \frac{1}
{2}A(M_\rho^2)   + \frac{1}
{2}A} \right) + P^{\mu \nu } \frac{{13}}
{6}B} \right] \nonumber\\ 
 &  = g^2 N_f \delta ^{ab} \left[ {g^{\mu \nu } \frac{3}
{4}M_\rho  ^2 B + P^{\mu \nu } \frac{{13}}
{6}B} \right], 
\end{align}
where we take $D\to4$ and use
\begin{align}
\lim_{D\to4}\frac{1}{D}(1 - \frac{1}{2}D)A(M_\rho^2)=\frac{1}{4}M_\rho^2B.
\end{align}
Similarly the divergent part including the internal line of $A_1$ is
\begin{align}
(5) + (14) + (16)
 &   = g^2 N_f \delta ^{ab} \left[ {g^{\mu \nu } \frac{3}
{4}M_{A_1 } ^2 B + P^{\mu \nu } \frac{{13}}
{6}B} \right].
\end{align} 
The sum of divergent part is given by
\begin{align}
\left.\Pi_{\rho\rho}^{\mu\nu}\right|_\text{div}
 & = g^2 N_f \delta ^{ab} \left[g^{\mu \nu } \left( A\left( { - (a - \beta )\gamma ^2  + \frac{{ - 1}}{\beta }(a - \beta ) - \frac{1}
{4}a^2 (1 - \gamma ^2 )^2  - \frac{1}
{{4\beta ^2 }}(2\beta  - a)^2  - \frac{1}
{4} + \frac{{ - \gamma ^2 (a - \beta )^2 }}
{{2\beta }}} \right)\right.\right.\nonumber\\ 
 &~~~~~~~~~~~~~~~~~~  - \left.F_\pi  ^2 g^2 \frac{3}
{4}B\left( {(\beta  - a)^2 \frac{1}
{\beta }(\beta \gamma ^2  + 1) - (a + \beta )} \right)\right)\nonumber\\
&~~~~~~~~~~~~~~~~~~   + \left.P^{\mu \nu } B\left( {\frac{{13}}
{3} + \frac{1}
{{24}}\left( { - a^2 (1 - \gamma ^2 )^2  - \frac{1}
{{\beta ^2 }}(2\beta  - a)^2  - 1 - \frac{{2\gamma ^2 (a - \beta )^2 }}
{\beta } - \frac{6}
{\beta }(\beta \gamma  - a)^2 } \right)} \right)\right]. 
\end{align}
The divergent part is canceled out by the following counterterms:
\begin{align}
\left. {\Pi _{\rho \rho }^{\mu \nu } } \right|_{{\text{div}}}  =  - (M_\rho  ^2 (\delta _{Z_g }  + \delta _{Z_\rho  } ) + g^2 \delta _{F_\sigma  ^2 } )g^{\mu \nu }  + (\delta _{Z_\rho  }  + 2\delta _\kappa  )P^{\mu \nu }.
\label{eq:cntrrhorho}
\end{align}
\subsection{$A_1-A_1$}
\begin{figure}
 \begin{align}
\Pi_{A_1A_1}^{\mu\nu}&=\vvs{$\pi$}+\vvs{$p$}+\vvs{$\sigma$}+\vvv{$\rho$}\nonumber\\
&+\vvv{$A_1$}+\vvss{$\sigma$}{$\pi$}+\vvss{$\sigma$}{$p$}+\vvvs{$A_1$}{$\sigma$}\nonumber\\
&+\vvvs{$\rho$}{$\pi$}+\vvvs{$\rho$}{$p$}+\vvvv{$A_1$}{$\rho$}+\vvgg{$C_A$}{$C_V$}.\nonumber
\end{align}
\caption{Contributions to $\Pi_{A_1A_1}^{\mu\nu}$ at one-loop. }
\label{fig:A_1A_1}
\end{figure}

The contributions from each diagram shown in Fig. \ref{fig:A_1A_1} are
 \begin{align}
(1):~~\scalebox{1.0}{\vvs{$\pi$}}
 &  = g^2 (a - \beta )\gamma ^2 N_f \delta ^{ab} g^{\mu \nu } A(0) ,\displaybreak[0]\\ 
(2):~~\scalebox{1.0}{\vvs{$p$}}
 &  = \frac{{g^2 }}
{\beta }(a - \beta )N_f \delta ^{ab} g^{\mu \nu } A(0) ,\displaybreak[0]\\ 
(3):~~\scalebox{1.0}{\vvs{$\sigma$}}
&=0,\displaybreak[0]\\ 
(4):~~\scalebox{1.0}{\vvv{$\rho$}}
 &  = g^2 N_f \delta ^{ab} g^{\mu \nu } \frac{{(D - 1)^2 }}
{D}A(M_\rho  ^2 ) ,\displaybreak[0]\\ \nonumber\\
(5):~~\scalebox{1.0}{\vvv{$A_1$}}
 &  = g^2 N_f \delta ^{ab} g^{\mu \nu } \frac{{(D - 1)^2 }}
{D}A(M_{A_1 } ^2 ) \displaybreak[0],\\ 
(6):\scalebox{1.0}{\vvss{$\sigma$}{$\pi$}}
 &  = \frac{{g^2 \gamma ^2 }}
{{4a}}N_f \delta ^{ab} ( (a - \beta )^2 B^{\mu \nu } (p,0,0) + (\beta \gamma  - a)^2 p^\mu  p^\nu  B(p,0,0) \nonumber\displaybreak[0]\\ 
 &~~~  + 2(\beta  - a)(\beta \gamma  - a)(p^\mu  B^\nu  (p,0,0) + p^\nu  B^\mu  (p,0,0) - p^\mu  p^\nu  B(p,0,0)))  ,\displaybreak[0]\\ 
(7):\scalebox{1.0}{\vvss{$\sigma$}{$p$}}
 &  = \frac{{g^2 }}
{{4a\beta }}N_f \delta ^{ab} ( a^2 B^{\mu \nu } (p,0,0) + (\beta \gamma  - a)^2 p^\mu  p^\nu  B(p,0,0) \nonumber\\ 
 &~~~  + 2a(a - \beta \gamma )(p^\mu  B^\nu  (p,0,0) + p^\nu  B^\mu  (p,0,0) - p^\mu  p^\nu  B(p,0,0))  ,\displaybreak[0]\\ 
(8):\scalebox{1.0}{\vvvs{$A_1$}{$\sigma$}}
&=0,\\ \nonumber\\
(9):\scalebox{1.0}{\vvvs{$\rho$}{$\pi$}}
&   =  - F_\pi  ^2 (\beta  - a)^2 \gamma ^2 g^4 N_f \delta ^{ab} \left( {g^{\mu \nu } B(p,0,M_\rho  ^2 ) - \frac{1}
{{M_\rho  ^2 }}\left( {\tilde B^{\mu \nu } (p,0,M_{\rho } ^2 ) - \tilde B^{\mu \nu } (p,0,0)} \right)} \right),\displaybreak[0]\\
(10):\scalebox{1.0}{\vvvs{$\rho$}{$p$}}
& = -\frac{{F_\pi  ^2 }}
{\beta }(\beta  - a)^2 g^4 N_f \delta ^{ab}\left( {g^{\mu \nu } B(p,0,M_\rho  ^2 ) - \frac{1}
{{M_\rho  ^2 }}\left( {\tilde B^{\mu \nu } (p,0,M_{\rho } ^2 ) - \tilde B^{\mu \nu } (p,0,0)} \right)} \right),\displaybreak[0]\\
(11):\vvvv{$A_1$}{$\rho$} 
 &  = g^2 N_f \delta ^{ab} {\int\frac{\textrm{d}^4k}{i(2\pi)^4}}\frac{1}
{{((k - p)^2  - M_\rho  ^2 )(k^2  - M_{A_1 } ^2 )}} \nonumber\\ 
 &~~~  \times \{  - 2p^\rho  g^{\mu \sigma }  + 2p^\sigma  g^{\mu \rho }  + (p - 2k)^\mu  g^{\rho \sigma } \} \{  - 2p^{\rho '} g^{\nu \sigma '}  + 2p^{\sigma '} g^{\nu \rho '}  + (p - 2k)^\nu  g^{\rho '\sigma '} \}  \nonumber\\ 
 &~~~  \times (g_{\rho \rho '}  - \frac{{k_\rho  k_{\rho '} }}
{{k^2 }})(g_{\sigma \sigma '}  - \frac{{(k - p)_\sigma  (k - p)_{\sigma '} }}
{{(k - p)^2 }}) ,\displaybreak[0]\\
(12):\scalebox{1.0}{\vvgg{$C_A$}{$C_V$}}
 &  = \frac{{ - g^2 }}
{2}N_f \delta ^{ab} (B^{\mu \nu } (p,0,0) - p^\mu  p^\nu  B(p,0,0)).
\end{align}
The divergent parts are given by
\begin{align}
 & (1):g^2 (a - \beta )\gamma ^2 N_f \delta ^{ab} g^{\mu \nu } A, \nonumber\displaybreak[0]\\ 
 & (2):\frac{{g^2 }}
{\beta }(a - \beta )N_f \delta ^{ab} g^{\mu \nu } A ,\nonumber\displaybreak[0]\\ 
 & (3):0 ,\nonumber\displaybreak[0]\\ 
 & (4):g^2 N_f \delta ^{ab} g^{\mu \nu } \frac{{(D - 1)^2 }}
{D}(A - M_\rho  ^2 B) ,\nonumber\displaybreak[0]\\ 
 & (5):g^2N_f \delta ^{ab} g^{\mu \nu } \frac{{(D - 1)^2 }}
{D}(A - M_{A_1 } ^2 B) ,\nonumber\displaybreak[0]\\ 
 & (6):\frac{{g^2 \gamma ^2 }}
{{4a}}N_f \delta ^{ab} \left( { - 2g^{\mu \nu } (a - \beta )^2 A - P^{\mu \nu } \frac{1}
{3}(a - \beta )^2 B + (\beta \gamma  - a)^2 p^\mu  p^\nu  B} \right),\nonumber\displaybreak[0]\\ 
 & (7):\frac{{g^2 }}
{{4a\beta }}N_f \delta ^{ab} \left( { - 2a^2 A g^{\mu \nu }  - P^{\mu \nu } \frac{{a^2 }}
{3}B + (\beta \gamma  - a)^2 p^\mu  p^\nu  } \right)   ,\nonumber\displaybreak[0]\\ 
 & (8):0 ,\nonumber\displaybreak[0]\\ 
 & (9): - N_f \delta ^{ab} F_\pi  ^2 (\beta  - a)^2 \gamma ^2 g^4 g^{\mu \nu } \frac{{D - 1}}
{D}B ,\nonumber\displaybreak[0]\\ 
 & (10): - N_f \delta ^{ab} \frac{{F_\pi  ^2 }}
{\beta }(\beta  - a)^2 g^4 g^{\mu \nu } \frac{{D - 1}}
{D}B ,\nonumber\displaybreak[0]\\ 
 & (11):N_f \delta ^{ab} g^2 \left( {P^{\mu \nu } \frac{{14}}
{3}B - 2g^{\mu \nu } A + 3g^{\mu \nu } (M_\rho  ^2  + M_{A_1 } ^2 )B - \frac{{g^{\mu \nu } p^2 }}
{2}B} \right)  ,\nonumber\displaybreak[0]\\ 
 & (12): - \frac{{g^2 }}
{2}N_f \delta ^{ab} \left( { - g^{\mu \nu } (2A + p^2 B) + P^{\mu \nu } \frac{{2B}}
{3}} \right). 
\end{align}
\begin{align}
 (4) + (5) + (11) + (12)
   = g^2 N_f \delta ^{ab} \left[ {g^{\mu \nu } \frac{3}
{4}(M_\rho  ^2  + M_{A_1 } ^2 )B + P^{\mu \nu } \frac{{13}}
{3}B} \right].
\end{align}
The sum of divergent parts is given by
\begin{align}
 \left. {\Pi _{A_1 A_1 }^{\mu \nu } } \right|_{{\text{div}}}  &= g^2 N_f \delta ^{ab} \left[g^{\mu \nu } \left(A\left( {(a - \beta )\gamma ^2  + \frac{1}
 {\beta }(a - \beta ) - \frac{{\gamma ^2 }}
{{2a}}(a - \beta )^2  - \frac{a}
{{2\beta }}} \right)\right.\right. \nonumber\\ 
 &~~~~~~~~~~~~~~~~~~  - \left.\frac{3}
{4} F_\pi  ^2 g^2 B( (\beta  - a)^2 \gamma ^2  + \frac{1}
{\beta }(\beta  - a)^2  - (a + \beta )) \right) \nonumber\\ 
 &~~~~~~~~~~~~~~~~~~  + \left.P^{\mu \nu } B\left( {\frac{{13}}
{3} - \frac{1}
{{12}}\frac{{\gamma ^2 }}
{a}(a - \beta )^2  + 3\frac{{\gamma ^2 }}
{a}(\beta \gamma  - a)^2  + \frac{1}
{{a\beta }}a^2  + 3\frac{1}
{{a\beta }}(\beta \gamma  - a)^2 } \right)\right].
\end{align}
The divergent part is canceled out by the following counterterms:
\begin{align}
\left. {\Pi _{A_1 A_1 }^{\mu \nu } } \right|_{{\text{div}}}  =  - (M_{A_1 } ^2 (\delta _{Z_g }  + \delta _{Z_{A_1 } } ) + g^2 \delta _{F_p ^2 } )g^{\mu \nu }  + (\delta _{Z_\rho  }  - 2\delta _\kappa  )P^{\mu \nu } .
\label{eq:cntrA_1A_1}
\end{align}
\subsection{$\pi-\pi$}
\begin{figure}
 \begin{align}
-\delta^{ab}\Pi_{\pi\pi}&= \scs{$\pi$}+\scs{$p$}+\scs{$\sigma$}+\scv{$\rho$}\nonumber\\
&+\scv{$A_1$}+\scss{$\sigma$}{$\pi$}+\scss{$p$}{$\sigma$}+\scvs{$\rho$}{$\pi$}\nonumber\\
&\scvs{$\rho$}{$p$}+\scvs{$A_1$}{$\sigma$}+\scvv{$\rho$}{$A_1$}.\nonumber
\end{align}
\caption{Contributions to $\Pi_{\pi\pi}$ at one-loop. }
\label{fig:pipi}
\end{figure}

The contributions from each diagram shown in Fig. \ref{fig:pipi} are
 \begin{align}
(1):\scalebox{1.0}{\scs{$\pi$}}
& =  - \frac{{4 - 3a(1 - \gamma ^2 )^2 }}{{12}}\frac{{p^2 }}
{{F_\pi  ^2 }}N_f \delta ^{ab} A(0)
, \\ 
(2):\scalebox{1.0}{\scs{$\sigma$} }
 &   = \frac{{\beta \gamma ^2 }}{a}\frac{{p^2 }}{{F_\pi  ^2 }}N_f \delta ^{ab} A(0)
, \\ 
(3):\scalebox{1.0}{\scs{$p$}}
 &   = \frac{{a\gamma ^2 }}{{4\beta }}\frac{{p^2 }}{{F_\pi  ^2 }}N_f \delta ^{ab} A(0)
, \\ 
(4):\scalebox{1.0}{\scv{$\rho$}}
 &  =  - g^2 (a - \beta )\gamma ^2 N_f \delta ^{ab} (1 - D)A(M_\rho  ^2 ), \\ 
(5):\scalebox{1.0}{\scv{$A_1$}}
 &= g^2 (a - \beta )\gamma ^2 \delta ^{ab} N_f (1 - 
  D)A(M_{A_1 } ^2 ),\displaybreak[0]\\ 
(6):\scalebox{1.0}{\scss{$\sigma$}{$\pi$} }
 &   = \frac{{a(1 - \gamma ^2 )^2 }}{4}\frac{{p^2 }}
{{F_\pi  ^2 }}N_f \delta ^{ab} (A(0) + p^2 B(p,0,0)), \displaybreak[0]\\ 
(7):\scalebox{1.0}{\scss{$p$}{$\sigma$}}
 & = \frac{{\gamma ^2 }}{{4a\beta }}\frac{{p^2 }}
{{F_\pi  ^2 }}N_f \delta ^{ab} (( - 4\beta a - 2\beta ^2  + a^2 )A(0) + (\beta  - a)^2 p^2 B(p,0,0))
,  \displaybreak[0]\\ 
(8):\scalebox{1.0}{\scvs{$\rho$}{$\pi$}}
 &   =  - g^2 a^2 (1 - \gamma ^2 )^2 N_f \delta ^{ab} \left( {p^2 B(p,0,M_\rho  ^2 ) - \frac{{p^\mu  p^\nu  }}
{{M_\rho  ^2 }}\left( {\tilde B_{\mu \nu } (p,0,M_\rho  ^2 ) - \tilde B_{\mu \nu } (p,0,0)} \right)} \right), \displaybreak[0]\\ 
(9):\scalebox{1.0}{\scvs{$\rho$}{$p$}}
 & = \frac{{ - 1}}
{\beta }g^2 (a - \beta )^2 \gamma ^2 N_f \delta ^{ab} \left( {p^2 B(p,0,M_\rho  ^2 ) - \frac{{p^\mu  p^\nu  }}
{{M_\rho  ^2 }}\left( {\tilde B_{\mu \nu } (p,0,M_\rho  ^2 ) - \tilde B_{\mu \nu } (p,0,0)} \right)} \right)
, \displaybreak[0]\\ 
(10):\scalebox{1.0}{\scvs{$A_1$}{$\sigma$}}
 &   = \frac{{ - \gamma ^2 }}{a}g^2 (\beta  - a)^2 N_f \delta ^{ab} \left( {p^2 B(p,0,M_{A_1 } ^2 ) - \frac{{p^\mu  p^\nu  }}
{{M_{A_1 } ^2 }}\left( {\tilde B_{\mu \nu } (p,0,M_{A_1 } ^2 ) - \tilde B_{\mu \nu } (p,0,0)} \right)} \right)
, \displaybreak[0]\\ 
(11):\scalebox{1.0}{\scvv{$\rho$}{$A_1$}}
&  = F_\pi  ^2 g^4 (\beta  - a)^2 \gamma ^2 N_f \delta ^{ab} \Big((D - 1)B(p,M_\rho  ^2 ,M_{A_1 } ^2 )- \frac{{p^2 }}
{{M_{A_1 } ^2 }}\left( {B(p,M_{A_1 } ^2 ,M_\rho  ^2 ) - B(p,0,M_\rho  ^2 )} \right)\nonumber\displaybreak[0]\\
&~~~ + \frac{{p^\mu  p^\nu  }}
{{M_{A_1 } ^2 M_\rho  ^2 }}(\tilde B_{\mu \nu } (p,M_{A_1 } ^2 ,M_\rho  ^2 ) - \tilde B_{\mu \nu } (p,0,M_\rho  ^2 ))\Big) .
\end{align}
The divergent parts are give by
\begin{align}
(1):& - \frac{{4 - 3a(1 - \gamma ^2 )^2 }}{{12}}\frac{{p^2 }}
{{F_\pi  ^2 }}N_f \delta ^{ab} A, \nonumber\displaybreak[0]\\ 
(2):&\frac{{\beta \gamma ^2 }}
{a}\frac{{p^2 }}{{F_\pi  ^2 }}N_f \delta ^{ab} A, \nonumber\displaybreak[0]\\ 
(3):&\frac{{\gamma ^2 }}{\beta }\frac{1}{4}a\frac{{p^2 }}
{{F_\pi  ^2 }}N_f \delta ^{ab} A, \nonumber\displaybreak[0]\\ 
 (4):&g^2 (a - \beta )\gamma ^2 N_f \delta ^{ab} (D - 1)(A - M_\rho  ^2 B), \nonumber\displaybreak[0]\\ 
 (5):& - g^2 (a - \beta )\gamma ^2 N_f \delta ^{ab} (D - 1)(A - M_{A_1 } ^2 B), \nonumber\displaybreak[0]\\ 
 (6):&\frac{{a(1 - \gamma ^2 )^2 }}
{4}\frac{{p^2 }}
{{F_\pi  ^2 }}N_f \delta ^{ab} (A + p^2 B), \nonumber\displaybreak[0]\\ 
(7):&\frac{{\gamma ^2 }}
{{4a\beta }}\frac{{p^2 }}
{{F_\pi  ^2 }}N_f \delta ^{ab} ((a^2  - 2\beta ^2  - 4a\beta )A + (\beta  - a)^2 p^2 B) , \nonumber\displaybreak[0]\\ 
(8):& - p^2 g^2 a^2 (1 - \gamma ^2 )^2 N_f \delta ^{ab} \frac{3}
{4}B, \nonumber\displaybreak[0]\\ 
(9):&\frac{{ - p^2 }}
{\beta }g^2 N_f \delta ^{ab} (a - \beta )^2 \gamma ^2 \frac{3}
{4}B, \nonumber\\ 
(10):&\frac{{ - p^2 \gamma ^2 }}
{a}g^2 N_f \delta ^{ab} (\beta  - a)^2 \frac{3}
{4}B, \nonumber\\ 
(11):&F_\pi  ^2 g^4(\beta  - a)^2 \gamma ^2 N_f \delta ^{ab} (D - 1)B.
\end{align} 
The sum of divergent parts is given by
\begin{align}
\left. {\Pi _{\pi \pi } } \right|_{{\text{div}}}  &=  - N_f \frac{{p^2 }}
{{F_\pi  ^2 }}\left[A\left( { - \frac{{4 - 3a(1 - \gamma ^2 )^2 }}
{{12}} + \frac{{\beta \gamma ^2 }}
{a} + \frac{{\gamma ^2 }}
{\beta }\frac{1}
{4}a + \frac{{a(1 - \gamma ^2 )^2 }}
{4} + \frac{{\gamma ^2 }}
{{4a\beta }}( - 4\beta a - 2\beta ^2  + a^2 )} \right)\right. \nonumber\\ 
 &~~~~~~~~~~~~~~~~\left.  - BF_\pi  ^2 g^2 \frac{3}
{4}\left( {a^2 (1 - \gamma ^2 )^2  + \gamma ^2 (a - \beta )^2 (\frac{1}
{\beta } + \frac{1}
{a})} \right) + \frac{1}
{4}p^2 B\left( {a(1 - \gamma ^2 )^2  + \frac{{\gamma ^2 }}
{{a\beta }}(\beta  - a)^2 } \right)\right] .
\end{align} 
The divergences are canceled out by the following counterterms:
\begin{align}
\left. {\Pi _{\pi \pi } } \right|_{{\text{div}}}  = p^2 (\delta _{Z_\pi  }  + p^2 \delta_{4\pi\pi} ),
\label{eq:cntrpipi}
\end{align}
where $\delta_{4\pi\pi}$ is the counter term from the ${\cal O}(p^4)$ Lagrangian.
\subsection{$\pi-p$}
\begin{figure}
 \begin{align}
\delta^{ab}\Pi_{\pi p}&= \scs{$\pi$}+\scs{$p$}+\scs{$\sigma$}+\scv{$\rho$}\nonumber\\
&+\scv{$A_1$}+\scss{$\sigma$}{$\pi$}+\scss{$p$}{$\sigma$}+\scvs{$\rho$}{$\pi$}\nonumber\\
&\scvs{$\rho$}{$p$}+\scvs{$A_1$}{$\sigma$}+\scvv{$\rho$}{$A_1$} .\nonumber
\end{align}
\caption{Contributions to $\Pi_{\pi p}$ at one-loop. }
\label{fig:pip}
\end{figure}

The contributions from each diagram shown in Fig. \ref{fig:pip} are
 \begin{align}
(1):\scalebox{1.0}{\scs{$\pi$}}
 &   =  - \frac{\gamma }{{2\sqrt \beta  }}\frac{{2\beta  - 3a}}{6}(1 - \gamma ^2 )\frac{{p^2 }}
{{F_\pi  ^2 }}N_f \delta ^{ab} A(0),\\ 
(2):\scalebox{1.0}{\scs{$\sigma$}}
 &  = \frac{\gamma }{{2\sqrt \beta  }}\frac{\beta }{a}\frac{{p^2 }}
{{F_\pi  ^2 }}N_f \delta ^{ab} A(0),\\ 
(3):\scalebox{1.0}{\scs{$p$}}
 &   = \frac{\gamma }
{{2\sqrt \beta  }}\frac{{a - 2\beta }}
{{6\beta }}\frac{{p^2 }}
{{F_\pi  ^2 }}N_f \delta ^{ab} A(0),\\ 
(4):\scalebox{1.0}{\scv{$\rho$}}
 &  = \frac{\gamma }
{{2\sqrt \beta  }}2g^2 (a - \beta )3N_f \delta ^{ab} A(M_\rho  ^2 ) ,\\ 
(5):\scalebox{1.0}{\scv{$A_1$}}
 &   = \frac{\gamma }{{2\sqrt \beta  }}2g^2 (\beta  - a)3N_f \delta ^{ab} A(M_{A_1 } ^2 ) \displaybreak[0],\\ 
(6):\scalebox{1.0}{\scss{$\sigma$}{$\pi$}}
 &   = \frac{\gamma }{{2\sqrt \beta  }}\frac{{ - 1}}{2}(1 - \gamma ^2 )\frac{{p^2 }}
{{F_\pi  ^2 }}N_f \delta ^{ab} ((a + 2\beta )A(0) + (a + \beta )p^2 B(p,0,0))  ,\displaybreak[0]\\ 
(7):\scalebox{1.0}{\scss{$p$}{$\sigma$}}
 & = \frac{\gamma }{{2\sqrt \beta  }}\frac{1}{{2\beta }}\frac{{p^2 }}
{{F_\pi  ^2 }}N_f \delta ^{ab} ((a - 2\beta )A(0) + (a - \beta )p^2 B(p,0,0)) ,\displaybreak[0]\\ 
(8):\scalebox{1.0}{\scvs{$\rho$}{$\pi$}}
 &   = \frac{\gamma }
{{2\sqrt \beta  }}2g^2 a(1 - \gamma ^2 )(a - \beta )N_f \delta ^{ab} \left( {p^2 B(p,0,M_\rho  ^2 ) - \frac{{p^\mu  p^\nu  }}
{{M_\rho  ^2 }}\left( {\tilde B_{\mu \nu } (p,0,M_\rho  ^2 ) - \tilde B_{\mu \nu } (p,0,0)} \right)} \right) ,\displaybreak[0]\\ 
(9):\scalebox{1.0}{\scvs{$\rho$}{$p$}}
&  = \frac{\gamma }
{{2\sqrt \beta  }}2g^2 \frac{{(2\beta  - a)}}
{\beta }(a - \beta )N_f \delta ^{ab} \left( {p^2 B(p,0,M_\rho  ^2 ) - \frac{{p^\mu  p^\nu  }}
{{M_\rho  ^2 }}\left( {\tilde B_{\mu \nu } (p,0,M_\rho  ^2 ) - \tilde B_{\mu \nu } (p,0,0)} \right)} \right) ,\displaybreak[0]\\ 
(10):\scalebox{1.0}{\scvs{$A_1$}{$\sigma$}}
 & = \frac{\gamma }
{{2\sqrt \beta  }}2g^2 (\beta  - a)N_f \delta ^{ab} \left( {p^2 B(p,0,M_\rho  ^2 ) - \frac{{p^\mu  p^\nu  }}
{{M_\rho  ^2 }}\left( {\tilde B_{\mu \nu } (p,0,M_\rho  ^2 ) - \tilde B_{\mu \nu } (p,0,0)} \right)} \right) ,\displaybreak[0]\\ 
(11):\scalebox{1.0}{\scvv{$\rho$}{$A_1$}}
 &  = \frac{\gamma }
{{2\sqrt \beta  }}2g^4 F_\pi  ^2 (a - \beta )^2 N_f \delta ^{ab} \Big(3B(p,M_\rho  ^2 ,M_{A_1 } ^2 ) - \frac{{p^2 }}
{{M_{A_1 } ^2 }}\left( {B(p,M_{A_1 } ^2 ,M_\rho  ^2 ) - B(p,0,M_\rho  ^2 )} \right) \nonumber\displaybreak[0]\\ 
 &~~~  + \frac{{p^\mu  p^\nu  }}
{{M_{A_1 } ^2 M_\rho  ^2 }}(\tilde B_{\mu \nu } (p,M_{A_1 } ^2 ,M_\rho  ^2 ) - \tilde B_{\mu \nu } (p,0,M_\rho  ^2 ))\Big).  
\end{align}
The divergent parts are given by
\begin{align}
(1):&\frac{\gamma }
{{2\sqrt \beta  }}\frac{{2\beta  - 3a}}
{6}(1 - \gamma ^2 )\frac{{p^2 }}
{{F_\pi  ^2 }}N_f \delta ^{ab} A, \nonumber\displaybreak[0]\\ 
(2):&\frac{\gamma }
{{2\sqrt \beta  }}\frac{\beta }
{a}\frac{{p^2 }}
{{F_\pi  ^2 }}N_f \delta ^{ab} A, \nonumber\displaybreak[0]\\ 
(3):&\frac{\gamma }
{{2\sqrt \beta  }}\frac{{a - 2\beta }}
{{6\beta }}\frac{{p^2 }}
{{F_\pi  ^2 }}N_f \delta ^{ab} A, \nonumber\displaybreak[0]\\ 
(4):&\frac{\gamma }
{{2\sqrt \beta  }}2g^2 (a - \beta )3N_f \delta ^{ab} (A - M_\rho  ^2 B), \nonumber\displaybreak[0]\\ 
(5):&\frac{\gamma }
{{2\sqrt \beta  }}2g^2 (\beta  - a)3N_f \delta ^{ab} (A - M_{A_1 } ^2 B), \nonumber\displaybreak[0]\\ 
(6):&\frac{\gamma }
{{2\sqrt \beta  }}\frac{{ - 1}}
{2}(1 - \gamma ^2 )\frac{{p^2 }}
{{F_\pi  ^2 }}N_f \delta ^{ab} ((a + 2\beta )A + (a + \beta )p^2 B), \nonumber\displaybreak[0]\\ 
(7):&\frac{\gamma }
{{2\sqrt \beta  }}\frac{1}
{{2\beta }}\frac{{p^2 }}
{{F_\pi  ^2 }}N_f \delta ^{ab} ((a - 2\beta )A + (a - \beta )p^2 B), \nonumber\\ 
(8):&\frac{\gamma }
{{2\sqrt \beta  }}g^2 a(1 - \gamma ^2 )(a - \beta )N_f \delta ^{ab} p^2 \frac{3}
{2}B, \nonumber\\ 
(9):&g^2 \frac{{(2\beta  - a)}}
{\beta }(a - \beta )N_f \delta ^{ab} p^2 \frac{3}
{2}B, \nonumber\\ 
(10):&\frac{{g^2 \gamma }}
{{2\sqrt \beta  }}(\beta  - a)N_f \delta ^{ab} p^2 \frac{3}
{2}B, \nonumber\\ 
(11):&\frac{\gamma }
{{2\sqrt \beta  }}2g^4 F_\pi  ^2 (a - \beta )^2 3N_f \delta ^{ab} B.
\end{align} 
The sum of divergent parts is given by
\begin{align}
\left. {\Pi _{\pi p} } \right|_{{\text{div}}}  &= \frac{\gamma }
{{2\sqrt \beta  }}\frac{{p^2 }}
{{F_\pi  ^2 }}N_f \delta ^{ab} \left[A\left( {(\frac{1}
{3}\beta  - \frac{1}
{2}a)(1 - \gamma ^2 ) + \frac{\beta }
{a} + \frac{{ - 1}}
{\beta }(\frac{1}
{3}\beta  - \frac{1}
{2}a) - (1 - \gamma ^2 )\frac{1}
{2}(a + 2\beta ) + \frac{1}
{{2\beta }}(a - 2\beta )} \right) \right.\nonumber\\ 
 &  ~~~~~~~~~~~~~~~~~~~~~~~~+ \left.2F_\pi  ^2 \frac{3}
{4}g^2 B\left( { - a(1 - \gamma ^2 )(\beta  - a) + \frac{{(2\beta  - a)}}
{\beta }(a - \beta ) + (\beta  - a)} \right)\right. \nonumber\\
&~~~~~~~~~~~~~~~~~~~~~~~~\left.+ \frac{{p^2 }}{2}B( - (1 - \gamma ^2 )(a + \beta ) + \frac{1}
{\beta }(a - \beta ))\right].
\end{align} 
The divergences are canceled out by the following counterterms:
\begin{align}
\left. {\Pi _{\pi p} } \right|_{{\text{div}}}  = p^2 ( - \delta _{\pi p} 
 + p^2 \delta_{4\pi p} ),
\label{eq:cntrpip}
\end{align}
where $\delta_{4\pi p}$ is the counterterm from ${\cal O}(p^4)$.
\section{Renormalization Group Equations}
The renormalization group equation for the parameter $\lambda_i=F_\pi^2,a,\beta, ...$,
at one loop is given by
\begin{align}
\mu \frac{d\lambda_i}{d\mu } = \left. \Lambda \frac{d\delta \lambda_i}{d\Lambda } \right|_{_\Lambda   = \mu } ,
\end{align}
where $\delta \lambda_i$ is counter term for  $\lambda_i$.
From Eqs.(\ref{eq:cntrAA}), (\ref{eq:cntrVV}), (\ref{eq:cntrAA_1}), 
(\ref{eq:cntrVrho}), (\ref{eq:cntrrhorho}), (\ref{eq:cntrA_1A_1}), 
(\ref{eq:cntrpipi}) and (\ref{eq:cntrpip}),
we obtain the renormalization group equations as follows:
\begin{align}
\mu \frac{{dF_\pi  ^2 }}{{d\mu }} &= \frac{{2N_f }}
{{(4\pi )^2 }}\left[\mu ^2 \left(1 - \frac{a}
{2} + \gamma ^2  + a\gamma ^2  - \frac{{a\gamma ^2 }}
{{2\beta }} - \frac{{\beta \gamma ^2 }}
{{2a}} - \frac{{a\gamma ^4 }}
{2}\right)\right. \nonumber\\ 
 &~~~~~~~~~~~~~  + \frac{3}
{4}F_\pi  ^2 g^2 \left.\left(a^2  - a\gamma ^2  - 2a^2 \gamma ^2  + \frac{{a^2 \gamma ^2 }}
{\beta } - \beta \gamma ^2  + \frac{{\beta ^2 \gamma ^2 }}
{a} + a^2 \gamma ^4 \right)\right], \nonumber\\ 
\mu\frac{dF_\sigma  ^2}{d\mu} & = \frac{2N_f}{(4\pi)^2} \left[\mu^2\left(\frac{1}
{4} + \frac{{a^2 }}{4} + \frac{{a^2 }}
{{4\beta ^2 }} - \frac{{a^2 \gamma ^2 }}
{2} + \frac{{a^2 \gamma ^2 }}
{{2\beta }} - \frac{{\beta \gamma ^2 }}
{2} + \frac{{a^2 \gamma ^4 }}
{4}\right)\right. \nonumber\\ 
 &~~~~~~~~~~~~~  + \frac{3}
{4}F_\pi  ^2g^2\left.\left( a  +\frac{{a^2 }}
{\beta } + a^2 \gamma ^2  + \beta ^2 \gamma ^2  - 2a\beta \gamma ^2 \right) \right], \nonumber\\ 
\mu\frac{dF_{A_1}  ^2}{d\mu}  &= \frac{2N_f}{(4\pi)^2} \gamma ^2 \left[\mu^2\left( - 1 - a + \frac{a}
{{2\beta }} + \frac{\beta }
{a} + \frac{{a\gamma ^2 }}
{2} + \frac{{\beta ^2 \gamma ^2 }}
{{2a}}\right)\right. \nonumber\\ 
 &~~~~~~~~~~~~~~~~~  + \frac{3}
{4}F_\pi  ^2 g^2\left.\left( a + 2a^2  - \frac{{a^2 }}
{\beta } - 2a\beta  - a^2 \gamma ^2  + \beta ^2 \gamma ^2  + 2\beta\right) \right],\nonumber\\
\mu \frac{{dF_p ^2 }}{{d\mu }} &= \frac{{2N_f }}
{{(4\pi )^2 }}\left[\mu ^2 \left( 1 - \frac{a}
{{2\beta }} - \frac{{a\gamma ^2 }}
{2} + \frac{{\beta ^2 \gamma ^2 }}
{{2a}}\right)\right. \nonumber\\ 
 &~~~~~~~~~~~~~  + \frac{3}
{4}F_\pi  ^2 g^2 \left.\left( - 3a + \frac{{a^2 }}
{\beta } + 4\beta  + a^2 \gamma ^2  - 2a\beta \gamma ^2  + \beta ^2 \gamma ^2 \right)\right], \nonumber\\
\mu\frac{dg^2}{d\mu} & = -2g^4\frac{N_f}{(4\pi)^2}\left[\frac{22}{3}-\frac{1}{48}\left(5+a^2+\frac{a^2}{b^2}-\frac{2a}{\beta}-2a\gamma^2-2a^2\gamma^2+\frac{2a^2\gamma^2}{\beta}-2\beta\gamma^2+\frac{2\beta^2\gamma^2}{a}+a^2\gamma^4\right)\right],\nonumber\\
\mu \frac{{dz_1  }}
{{d\mu }} &= \frac{{N_f }}
{{(4\pi )^2 }}\frac{1}
{{24}}\left(5 - 4a + a^2  + \frac{{a^2 }}
{{\beta ^2 }} - 2a^2 \gamma ^2  + \frac{{2a^2 \gamma ^2 }}
{\beta } + 2\beta \gamma ^2  + a^2 \gamma ^4 \right), \nonumber\\
\mu \frac{{dz_2  }}
{{d\mu }} &= \frac{{N_f }}
{{(4\pi )^2 }}\frac{1}
{{12}}\left(a - 2\beta \gamma ^2  + \frac{{\beta \gamma ^2 }}
{a} + \frac{{\beta ^2 \gamma ^4 }}
{a}\right), \nonumber\\ 
\mu \frac{{dz_3  }}
{{d\mu }} &= \frac{{N_f }}
{{(4\pi )^2 }}\frac{1}
{{12}}\left(1 + 2a - a^2  - \frac{{a^2 }}
{{\beta ^2 }} + \frac{{2a}}
{\beta } + 2a\gamma ^2  + 2a^2 \gamma ^2  - \frac{{2a^2 \gamma ^2 }}
{\beta } - 2\beta \gamma ^2  - a^2 \gamma ^4 \right) \nonumber\\ 
\mu \frac{{dz_4  }}
{{d\mu }} &= \frac{{N_f }}
{{(4\pi )^2 }}\frac{1}
{6}\gamma \left(1 - a + \beta  + \beta \gamma ^2  - \frac{{\beta ^2 \gamma ^2 }}
{a}\right),\nonumber\\
\mu \frac{{d\kappa }}{{d\mu }} &= \frac{{N_f }}
{{(4\pi )^2 }}\frac{{g^2 }}{{48}}\left( { - 5 - a^2  - \frac{{a^2 }}
{{\beta ^2 }} + \frac{{6a}}{\beta } + 6a\gamma ^2  + 2a^2 \gamma ^2  - \frac{{2a^2 \gamma ^2 }}
{\beta } - 6\beta \gamma ^2  + \frac{{2\beta ^2 \gamma ^2 }}
{a} - a^2 \gamma ^4 } \right)
\label{RGE1}
\end{align}

We define the parameter to calculate the RGEs for the dimensionless 
parameter for convenience:
\begin{align}
x(\mu ) &= \frac{{N_f }}
{{(4\pi )^2 }}\frac{{\mu ^{2} }}
{{F_\pi  ^2 }},\nonumber\\
G(\mu) &=  \frac{{N_f }}
{{(4\pi )^2 }}g^2.
\end{align}
The beta functions for the dimensionless parameters
$x,a,\beta,\gamma^2,G$ are given by
\begin{align}
\beta _x  &= x(2 - \frac{1}
{{F_\pi  ^2 }}\mu \frac{{dF_\pi  ^2 }}
{{d\mu }}) ,\nonumber\\ 
\beta _a  &= \frac{1}
{{F_\pi  ^2 }}(\mu \frac{{dF_\sigma  ^2 }}
{{d\mu }} - \mu \frac{{dF_\pi  ^2 }}
{{d\mu }}a) ,\nonumber\\ 
\beta _\beta   &= \frac{1}
{{F_\pi  ^2 }}(\mu \frac{{dF_p ^2 }}
{{d\mu }} - \mu \frac{{dF_\pi  ^2 }}
{{d\mu }}\beta ), \nonumber\\ 
\beta _{\gamma^2 }  &= \frac{1}
{{\beta F_\pi  ^2 }}(\mu \frac{{dF_{A_1 } ^2 }}
{{d\mu }} - \mu \frac{{dF_p ^2 }}
{{d\mu }}\gamma ^2 ),\nonumber\\ 
\beta_G &=\frac{{N_f }}{{(4\pi )^2 }}\mu\frac{d g^2}{d\mu}.
\label{RGE2}
\end{align}
We obtain the RGEs in Eq.(\ref{RGEs}) by combining Eqs.(\ref{RGE1}) and (\ref{RGE2}).

\end{document}